\documentclass[prb,12pt]{revtex4}
\usepackage{amsmath,amsthm,amssymb}
\usepackage{latexsym}
\usepackage{array}
\usepackage{amscd,graphicx}
\usepackage{bm,textcomp}

\begin{document}

\title{Current in a spin-orbit-coupling system}
\author{Yi Li}
\affiliation{Department of Physics, Fudan University, Shanghai 200433, China}
\author{Ruibao Tao}
\thanks{ To whom correspondence should be addressed. Email:
rbtao@fudan.edu.cn}
\affiliation{Chinese Center of Advanced Science and Technology (World Laboratory) , \\
P. O. Box 8730 Beijing 100080, China\\
Department of Physics, Fudan University, Shanghai 200433, China}
\date{\today}

\begin{abstract}
The formulae of particle current as well as spin- and angular momentum
currents are studied for most spin-orbit coupling (SOC) systems. It is shown
that the conventional expression of currents in some literatures are not
complete for some SOC systems. The particle current in Dresselhaus system
must have extra terms in additional to the conventional one, but no extra
term for Rashba, Luttinger model. Further more, we also prove that the extra
terms of total angular momentum appear in Rashba current in addition to
conventional one.
\end{abstract}

\maketitle

\section{Introduction}

In quantum mechanics calculating current is crucial to applications. A
widely accepted approach is using the correspondence regulation from
classical to quantum mechanics. For example, the formula of charge current
density is introduced as%
\begin{equation*}
\overrightarrow {j}(\overrightarrow{r},t)=e{Re}\{\sum\limits_{n}\rho
_{n}\Psi _{n}^{\dag }(\overrightarrow{r},t)\widehat{\overrightarrow{v}}\Psi
_{n}(\overrightarrow{r},t)\},
\end{equation*}%
where $e$ is an electron (or hole) charge, $\widehat{\overrightarrow{v}}$
the velocity operator defined by $\widehat{\overrightarrow{v}}=\frac{1}{%
i\hbar }[\overrightarrow{r},\widehat{H}],$ $\widehat{H}$ is the
Hamiltonian,\ $\Psi _{n}(\overrightarrow{r},t)$\ is the wave function of an
admissible state $|\Psi _{n}\rangle $\ of the Hamiltonian system and $\rho
_{n}$ is the probability of $\Psi _{n}(\overrightarrow{r},t)$ appeared in
the quantum mixed state. For example, if the system is in equilibrium, the
distribution of the probability is $\rho _{n}=e^{-\beta E_{n}},$ where $%
\beta =1/(kT),$ $E_{n}$ the energy of state $\Psi _{n}(\overrightarrow{r}%
,t). $\ Summation $n$ is taken to all admissible states of Hamiltonian.
Further, the definition of charge current density is extended to define some
other currents like spin current density%
\begin{equation*}
j_{y}^{z}=\frac{1}{2}{Re}\{\sum\limits_{n}\rho _{n}\Psi _{n}^{\dag }(%
\overrightarrow{r},t)(\widehat{s}^{z}\widehat{v}_{y}+\widehat{v}_{y}\widehat{%
s}^{z})\Psi _{n}(\overrightarrow{r},t)\}
\end{equation*}%
and orbital angular momentum current density%
\begin{equation*}
\overrightarrow{J^{z}}=\frac{1}{2}{Re}\left\{ \sum\limits_{n}\rho _{n}\Psi
_{n}^{\dagger }(\overrightarrow{r},t)\left[ \widehat{l}^{z}\widehat{%
\overrightarrow{v}}+\widehat{\overrightarrow{v}}\widehat{l}^{z}\right] \Psi
_{n}(\overrightarrow{r},t)\right\}
\end{equation*}%
where $l^{z}=(\overrightarrow{r}\times \overrightarrow{p})_{z}$\ is orbital
angular momentum\ along $z$ direction. And symmetrization in the order of $%
\widehat{\overrightarrow{v}}$ and $\widehat{s}^{z}$ (or $\widehat{l}^{z}$)
has been applied. The above conventional formula have been widely accepted
as natural definition of currents in literatures, especially in the ones of
semiconductor systems including spin-orbit-coupling (SOC).

In this paper, we will prove that the conventional formula of particle
current are not generally correct. They can not satisfy the continuity
equation for some systems. Based on the continuity equation,[9] we start our
calculation from $\partial w_{A}(\overrightarrow{r},t)/\partial t$\ for
observable quantity $A,$\ as will be illustrated in Sec.2, and utilize the
continuity equation to obtain the definition of $A$-current $\overrightarrow{%
j}_{A}$. We can find in the case like 2d Rashba, 3d Luttinger and the spin
independent Hamiltonian $\widehat{H}=\frac{1}{2m^{\ast }}\widehat{p}^{2}+V(%
\overrightarrow{r})$, where $V(\overrightarrow{r})$ can be an arbitrary
position dependent potential, the conventional particle current formula are
still corrected. However, we prove that the particle current will have a
non-trivial additional term except conventional one in some semiconductor,
such as the system with Dresselhaus SOC and other model Hamiltonians
including the terms $\overrightarrow{p}^{n}$ of momentum operator with power
$n>2$. We will derive the expressions of the extra term for these specific
systems.

This paper will focus on some systems with spin-orbit-coupling such as 2d
Rashba[1], 3d Luttinger[2] as well as Dresselhaus systems[3] that attracted
great interest since the experimental demonstration [4-6] of spin hall
effect in some semiconductor materials. Our study in this letter is based on
non-relativistic quantum mechanics.

\section{Particle current for some complex systems}

In general, we study an observable quantity $A$. From quantum mechanics, $A$
corresponds to a Hermitian operator $\widehat{A},$ then
\begin{eqnarray*}
\left\langle A\right\rangle &=&Tr\{\widehat{A}\widehat{\rho }\}=Tr\sum_{n}\{%
\widehat{A}|\Psi _{n}\rangle \rho _{n}\langle \Psi _{n}|\} \\
&=&\sum_{n}\rho _{n}\langle \Psi _{n}|\widehat{A}|\Psi _{n}\rangle
=\sum_{n}\rho _{n}\int d\overrightarrow{r}\langle \Psi _{n}|\overrightarrow{r%
}\rangle \langle \overrightarrow{r}|\widehat{A}|\Psi _{n}\rangle \\
&=&\sum_{n}\rho _{n}\int d\overrightarrow{r}\Psi _{n}^{\dagger }\left(
\overrightarrow{r},t\right) \widehat{A}\Psi _{n}\left( \overrightarrow{r}%
,t\right) ,
\end{eqnarray*}%
where we have used $\langle \overrightarrow{r}|\widehat{A}|\overrightarrow{r}%
^{\prime }\rangle =\widehat{A}(\overrightarrow{r}^{\prime })\delta (%
\overrightarrow{r}-\overrightarrow{r}^{\prime })$ which is diagonal in
position representation for the most of cases in quantum mechanics$,$ $%
\widehat{A}(\overrightarrow{r}^{\prime })$ is the operator $\widehat{A}$ in
position representation and simply expressed by $\widehat{A}.$ $\widehat{%
\rho }(=\sum_{n}|\Psi _{n}\rangle \rho _{n}\langle \Psi _{n}|)$ is the
density matrix of a mixed state, $\rho _{n}$ ($\in \lbrack 0,1]$) the
probability of $|\Psi _{n}\rangle $ appeared in the mixed state and $%
\sum\limits_{n}\rho _{n}=1$. $|\Psi _{n}\rangle $ is assumed here to satisfy
the time dependent Schr\"{o}dinger equation $i\hbar \partial (|\Psi
_{n}\rangle )/\partial t=\widehat{H}|\Psi _{n}\rangle $. Since $\left\langle
A\right\rangle $ is real, so the above equation is equivalent to
\begin{equation}
\left\langle A\right\rangle \equiv {Re}\sum_{n}\rho _{n}\int d%
\overrightarrow{r}\langle \Psi _{n}|\overrightarrow{r}\rangle \langle
\overrightarrow{r}|\widehat{A}|\Psi _{n}\rangle =\int w_{A}(\overrightarrow{r%
},t)d\overrightarrow{r}  \label{1}
\end{equation}%
and
\begin{eqnarray*}
w_{A}(\overrightarrow{r},t) &=&\sum_{n}\rho _{n}{Re}\langle \Psi _{n}|%
\overrightarrow{r}\rangle \langle \overrightarrow{r}|\widehat{A}|\Psi
_{n}\rangle \\
&=&\sum_{n}\rho _{n}{Re}\left\{ \Psi _{n}^{\dag }(\overrightarrow{r},t)%
\widehat{A}\Psi _{n}(\overrightarrow{r},t)\right\} ,
\end{eqnarray*}%
\begin{eqnarray*}
w_{A}(\overrightarrow{r},t) &=&\sum_{n}\rho _{n}w_{A}^{(n)}(\overrightarrow{r%
},t), \\
w_{A}^{(n)}(\overrightarrow{r},t) &=&{Re}\left\{ \Psi _{n}^{\dag }(%
\overrightarrow{r},t)\widehat{A}\Psi _{n}(\overrightarrow{r},t)\right\} .
\end{eqnarray*}%
Where $w_{A}(\overrightarrow{r},t)$ is defined as the density of quantity $%
\langle A\rangle $ at time $t$ and position $\overrightarrow{r}.$ When $[%
\widehat{A},\widehat{H}]=0,$
\begin{eqnarray*}
\frac{\partial \left\langle A\right\rangle }{\partial t} &=&\frac{\partial }{%
\partial t}\sum_{n}\rho _{n}{Re}\langle \Psi _{n}|\widehat{A}|\Psi
_{n}\rangle \\
&=&\frac{1}{i\hbar }\sum_{n}\rho _{n}{Re}\langle \Psi _{n}|\widehat{A}%
\widehat{H}-\widehat{H}\widehat{A}|\Psi _{n}\rangle =0
\end{eqnarray*}%
$A$ is a conserved quantity, therefore it yields a continuity equation:
\begin{equation*}
\frac{\partial w_{A}(\overrightarrow{r},t)}{\partial t}=-\nabla \cdot
\overrightarrow{j}_{A}.
\end{equation*}%
Particle number is one of conserved quantity, here it is relevant to choose $%
A=NI,$ $I$ is an identity operator. In general, one define a conserved
current related to mechanical quantity $A$ from above continuity equation.
Now the key point is to find out the calculation formula of current $%
\overrightarrow{j}_{A}$ for each specific Hamiltonian. When $A$ is relevant
to the total particle number $N,A=NI,I$ is identity operator, it yields a
definition of particle current $\overrightarrow{j}$. Total number of charges
$eN$ is also a conserved quantity, the relevant operator $A=eNI$ and
conserved charge current $\overrightarrow{j}_{e}=e\overrightarrow{j}$. In
general, $A$ could be others, so one can define other currents. In this
note, we only consider mixed state with time independent probability $\rho
_{n}$. $A$ is also studied within time independent.

For the particle with spin $s=1/2$, the state $\left\vert \Psi \right\rangle
$ can be described by
\begin{eqnarray*}
|\Psi \rangle &=&\left(
\begin{array}{c}
|\psi _{1}\rangle \\
|\psi _{2}\rangle%
\end{array}%
\right) , \\
\Psi (\overrightarrow{r},t) &=&\langle \overrightarrow{r}|\Psi \rangle
=\left(
\begin{array}{c}
\langle \overrightarrow{r}|\psi _{1}\rangle \\
\langle \overrightarrow{r}|\psi _{2}\rangle%
\end{array}%
\right) =\left(
\begin{array}{c}
\psi _{1}(\overrightarrow{r},t) \\
\psi _{2}(\overrightarrow{r},t)%
\end{array}%
\right) , \\
\Psi ^{\dag }(\overrightarrow{r},t) &=&\langle \Psi |\overrightarrow{r}%
\rangle =\left(
\begin{array}{cc}
\psi _{1}^{\ast }(\overrightarrow{r},t) & \psi _{2}^{\ast }(\overrightarrow{r%
},t)%
\end{array}%
\right) ,
\end{eqnarray*}%
where $\psi _{i}^{\ast }(\overrightarrow{r},t)$ is complex conjugation of $%
\psi _{i}(\overrightarrow{r},t)$ $(i=1,2).$\ The Schr\"{o}dinger (or Pauli )
equation is%
\begin{equation*}
\frac{\partial }{\partial t}\left\vert \Psi \right\rangle =\frac{1}{i\hbar }%
\widehat{H}\left\vert \Psi \right\rangle ,
\end{equation*}%
where Hamiltonian $\widehat{H}$ is a $2\times 2$ matrix. Since now, for
simplifying notation, we will use $\Psi $ to denote $\Psi (\overrightarrow{r}%
,t),\Psi ^{\dag }$ to $\Psi ^{\dag }(\overrightarrow{r},t).$ The time
evolution of quantity $w_{A}^{(n)}(\overrightarrow{r},t)$ is
\begin{equation}
\frac{\partial w_{A}^{(n)}(\overrightarrow{r},t)}{\partial t}={Re}\left\{
\left( \frac{\partial }{\partial t}\Psi _{n}^{\dag }\right) \widehat{A}\Psi
_{n}+\Psi _{n}^{\dag }\widehat{A}\frac{\partial }{\partial t}\Psi
_{n}\right\}  \label{2}
\end{equation}%
In equation (2), the operator $\widehat{A}$ has been represented in position
space. Meanwhile, it must bear in mind that $\Psi _{n}$ and $\Psi _{n}^{\dag
}$ are the time and position dependent wave function. When $\widehat{A}=%
\widehat{I}_{2}$ (a $2\times 2$ identity matrix),%
\begin{eqnarray*}
w(\overrightarrow{r},t) &=&n(\overrightarrow{r},t)={Re}\sum_{n}\left\langle
\overrightarrow{r}\left\vert \Psi _{n}\right\rangle \rho _{n}\left\langle
\Psi _{n}\right\vert \overrightarrow{r}\right\rangle \\
&\equiv &\sum_{n}\rho _{n}\Psi _{n}^{\dag }\Psi _{n},
\end{eqnarray*}%
$n(\overrightarrow{r},t)$ is the particle density, so the following
continuity equation defines the particle current density $\overrightarrow{j}%
. $%
\begin{equation*}
\frac{\partial n(\overrightarrow{r},t)}{\partial t}=-\nabla \cdot
\overrightarrow{j}(\overrightarrow{r},t).
\end{equation*}%
It describes the conservation of particle number. For a time independent
Hamiltonian, the wave function $\Psi _{n}(t)=\exp (-iE_{n}t/\hbar )\Psi
_{n}(0)$, $\widehat{H}\Psi _{n}(0)=E_{n}\Psi _{n}(0)$. Therefore we have $%
\partial n(r,t)/\partial t=0.$ It reduces to $\nabla \cdot \overrightarrow{j}%
(\overrightarrow{r},t)=0.$

For many Hamiltonians without spin-orbit-coupling, $\widehat{H}=\widehat{H}%
_{0}+V(\overrightarrow{r}),$ $\widehat{H}_{0}=\widehat{p}^{2}/2m^{\ast
}=-(\hbar ^{2}/2m^{\ast })\nabla ^{2}$. In terms of the equality $\widehat{H}%
_{0}=-$ $\frac{1}{2}\nabla \cdot \lbrack \overrightarrow{r},\widehat{H}%
_{0}]=-\frac{1}{2}\nabla \cdot \lbrack \overrightarrow{r},\widehat{H}],$
equation (1) can be easily calculated as
\begin{eqnarray}
\frac{\partial w_{A}^{(n)}(\overrightarrow{r},t)}{\partial t} &=&{Re}\left\{
\left( \frac{1}{i\hbar }\widehat{H}_{0}\Psi _{n}\right) ^{\dag }\Psi
_{n}+\Psi _{n}^{\dag }\frac{1}{i\hbar }\widehat{H}_{0}\Psi _{n}\right\}
\notag \\
&&+{Re}\left\{ \left( \frac{1}{i\hbar }V(\overrightarrow{r})\Psi _{n}\right)
^{\dag }\Psi _{n}+\Psi _{n}^{\dag }\frac{1}{i\hbar }V(\overrightarrow{r}%
)\Psi _{n}\right\}  \notag \\
&=&-\frac{1}{2}{Re}\left\{ \left( \nabla \cdot \frac{1}{i\hbar }[%
\overrightarrow{r},\widehat{H}_{0}]\Psi _{n}\right) ^{\dag }\Psi _{n}+\Psi
_{n}^{\dag }(\nabla \cdot \frac{1}{i\hbar }[\overrightarrow{r},\widehat{H}%
_{0}]\Psi _{n})\right\}  \notag \\
&=&-\frac{1}{2}{Re}\nabla \cdot \left\{ \left( \frac{1}{i\hbar }[%
\overrightarrow{r},\widehat{H}_{0}]\Psi _{n}\right) ^{\dag }\Psi _{n}+\Psi
_{n}^{\dag }(\frac{1}{i\hbar }[\overrightarrow{r},\widehat{H}_{0}]\Psi
_{n})\right\}  \notag \\
&=&-\nabla \cdot {Re}\left\{ \Psi _{n}^{\dag }(\frac{1}{i\hbar }[%
\overrightarrow{r},\widehat{H}_{0}]\Psi _{n}\right\} \equiv -\nabla \cdot {Re%
}\left\{ \Psi _{n}^{\dag }(\frac{1}{i\hbar }[\overrightarrow{r},\widehat{H}%
]\Psi _{n}\right\}  \label{3}
\end{eqnarray}%
In equation (3), $\ $we have used the real property of potential $V(%
\overrightarrow{r})$ that results $\left( \frac{1}{i\hbar }V(\overrightarrow{%
r})\Psi _{n}\right) ^{\dag }\Psi _{n}+\Psi _{n}^{\dag }\frac{1}{i\hbar }V(%
\overrightarrow{r})\Psi _{n}=0.$ Thus we obtain the standard continuity
equation and the formula of particle current
\begin{eqnarray}
\frac{\partial n(\overrightarrow{r},t)}{\partial t} &=&-\nabla \cdot {Re}%
\left\{ \sum\limits_{n}\rho _{n}\Psi _{n}^{\dag }\{\frac{1}{i\hbar }[%
\overrightarrow{r},\widehat{H}]\Psi _{n}\}\right\} ,  \notag \\
\overrightarrow{j}(\overrightarrow{r},t) &=&{Re}\left\{ \sum\limits_{n}\rho
_{n}\Psi _{n}^{\dag }\left( \widehat{\overrightarrow{v}}\Psi _{n}\right)
\right\} .  \label{4}
\end{eqnarray}%
Equation (4) is called as "conventional formula". The formula is general for
any potential $V(\overrightarrow{r})$ in Hamiltonian $\widehat{H}$ if the
potential is position dependent only. However, if potential $V$ contains
momentum operator like the case appeared in some effective Hamiltonians of
semiconductors, it will be different and the above current formula may need
to be confirmed or to be modified to including some extra terms. In recent
literatures, ones usually extend it to define other kinds of conventional
current like spin current, $\widehat{A}=\widehat{\overrightarrow{S}}=\frac{%
\hbar }{2}\widehat{\overrightarrow{\sigma }},$
\begin{equation}
\overrightarrow{j}_{\overrightarrow{S}}(\overrightarrow{r},t)=\frac{1}{2}{Re}%
\left\{ \sum\limits_{n}\rho _{n}\Psi _{n}^{\dag }\left( \widehat{%
\overrightarrow{v}}\widehat{\overrightarrow{S}}+\widehat{\overrightarrow{v}}%
\widehat{\overrightarrow{S}}\right) \Psi _{n})\right\} .  \label{5}
\end{equation}%
, but not conserved, where symmetric presentation is applied for two
non-commutating operators $\widehat{\overrightarrow{v}}$ and $\widehat{%
\overrightarrow{S}}$.

Now we discuss the particle current in some complex systems with SOC in
which the momentum operator appears in "potential" of Hamiltonian. First, we
study the current formula for the Dresselhaus Hamiltonian that contains
triple power of momentum operator. The Hamiltonian is
\begin{eqnarray*}
\widehat{H} &=&\widehat{H}_{0}+\widehat{H}_{D} \\
\widehat{H}_{0} &=&\frac{1}{2m^{\ast }}\overrightarrow{p}^{2}+V(%
\overrightarrow{r}) \\
\widehat{H}_{D} &=&\eta \lbrack p_{x}\left( p_{y}^{2}-p_{z}^{2}\right)
\sigma _{x}+p_{y}\left( p_{z}^{2}-p_{x}^{2}\right) \sigma _{y}+p_{z}\left(
p_{x}^{2}-p_{y}^{2}\right) \sigma _{z}],
\end{eqnarray*}%
where $\{\sigma _{i}:i=x,y,z\}$ are Pauli matrices. The question is whether
the particle current can still be calculated by conventional formula (4). We
rewrite the Hamiltonian as
\begin{equation*}
\widehat{H}=\widehat{H}_{0}+\widehat{H}_{D}=-\frac{1}{2}\nabla \cdot \lbrack
\overrightarrow{r},\widehat{H}_{0}]+V(\overrightarrow{r})+\widehat{H}_{D},
\end{equation*}%
It yields
\begin{eqnarray*}
\frac{\partial n(\overrightarrow{r},t)}{\partial t} &=&{Re}%
\sum\limits_{n}\rho _{n}\left[ \Psi _{n}^{\dagger }\left( \frac{1}{i\hbar }%
\widehat{H}\Psi _{n}\right) +\left( \frac{1}{i\hbar }\widehat{H}\Psi
_{n}\right) ^{\dagger }\Psi _{n}\right] \\
&=&{Re}\sum\limits_{n}\rho _{n}\left[ \Psi _{n}^{\dagger }\left( \frac{1}{%
i\hbar }\widehat{H}_{0}\Psi _{n}\right) +\left( \frac{1}{i\hbar }\widehat{H}%
_{0}\Psi _{n}\right) ^{\dagger }\Psi _{n}\right] \\
&&+{Re}\sum\limits_{n}\rho _{n}\left[ \Psi _{n}^{\dagger }\left( \frac{1}{%
i\hbar }\widehat{H}_{D}\Psi _{n}\right) +\left( \frac{1}{i\hbar }\widehat{H}%
_{D}\Psi _{n}\right) ^{\dagger }\Psi _{n}\right] .
\end{eqnarray*}%
Similar to the deduction of equation (3) for the terms having $\widehat{H}%
_{0}$ in right side of above equation, we directly have%
\begin{eqnarray}
\frac{\partial n(\overrightarrow{r},t)}{\partial t} &=&-\nabla \cdot {Re}%
\sum\limits_{n}\rho _{n}\left[ \Psi _{n}^{\dagger }\left( \frac{1}{i\hbar }%
[r,\widehat{H}_{0}]\Psi _{n}\right) \right]  \notag \\
&&+{Re}\sum\limits_{n}\rho _{n}\left[ \Psi _{n}^{\dagger }\left( \frac{1}{%
i\hbar }\widehat{H}_{D}\Psi _{n}\right) +\left( \frac{1}{i\hbar }\widehat{H}%
_{D}\Psi _{n}\right) ^{\dagger }\Psi _{n}\right] .  \label{6}
\end{eqnarray}%
If the second term in right side of above equation equals to $-\nabla \cdot {%
Re}\sum\limits_{n}\rho _{n}\left[ \Psi _{n}^{\dagger }\left( \frac{1}{i\hbar
}[r,\widehat{H}_{D}]\Psi _{n}\right) \right] ,$ the final formula of
particle current is just the conventional one. But, in deed, they are
different. Appendix $(A)$ gives the detail proof to show that an extra term
must exist in addition to the conventional current formula. The continuity
equation should be
\begin{eqnarray}
\frac{\partial n(\overrightarrow{r},t)}{\partial t} &=&-\nabla \cdot (%
\overrightarrow{j}_{conv}+\overrightarrow{j}_{extra}),  \notag \\
\overrightarrow{j}_{conv} &=&{Re}\{\sum\limits_{n}\rho _{n}\Psi _{n}^{\dag }(%
\frac{1}{i\hbar }[\overrightarrow{r},\widehat{H}]\Psi _{n})\},  \notag \\
\overrightarrow{j}_{extra} &=&{Re}\sum\limits_{n}\rho _{n}\overrightarrow{j}%
_{extra}^{\left( n\right) }  \label{7}
\end{eqnarray}%
\begin{eqnarray*}
(\overrightarrow{j}_{extra}^{\left( n\right) })_{x} &=&\frac{1}{6}\hbar
^{2}\eta \{(\nabla _{y}^{2}-\nabla _{z}^{2})[\Psi _{n}^{\dag }(\sigma
_{x}\Psi _{n})] \\
&&-2\nabla _{x}\nabla _{y}[\Psi _{n}^{\dag }(\sigma _{y}\Psi _{n})]+2\nabla
_{x}\nabla _{z}[\Psi _{n}^{\dag }\left( \sigma _{z}\Psi _{n}\right) ])\} \\
(\overrightarrow{j}_{extra}^{\left( n\right) })_{y} &=&\frac{1}{6}\hbar
^{2}\eta \{(\nabla _{z}^{2}-\nabla _{x}^{2})[\Psi _{n}^{\dag }(\sigma
_{y}\Psi _{n})] \\
&&-2\nabla _{y}\nabla _{z}[\Psi _{n}^{\dag }(\sigma _{z}\Psi _{n})]+2\nabla
_{y}\nabla _{x}[\Psi _{n}^{\dag }\left( \sigma _{x}\Psi _{n}\right) ])\} \\
(\overrightarrow{j}_{extra}^{\left( n\right) })_{z} &=&\frac{1}{6}\hbar
^{2}\eta \{(\nabla _{x}^{2}-\nabla _{y}^{2})[\Psi _{n}^{\dag }(\sigma
_{z}\Psi _{n})] \\
&&-2\nabla _{z}\nabla _{x}[\Psi _{n}^{\dag }(\sigma _{x}\Psi _{n})]+2\nabla
_{z}\nabla _{y}[\Psi _{n}^{\dag }\left( \sigma _{y}\Psi _{n}\right) ])\}
\end{eqnarray*}%
where $\overrightarrow{j}_{conv}$ is the conventional formula of particle
current density. In appendix A, we also prove that the term $\overrightarrow{%
j}_{extra}$ is real and position dependent if the position dependent
potential $V(\overrightarrow{r})$ is included in Hamiltonian. Therefore, $%
\nabla \cdot \overrightarrow{j}_{extra}\neq 0,$ so the nontrivial extra term
of particle current do appear in Dresselhaus system.

We also checked the formulae of particle current for 2d Rashba\textrm{\ }$%
\left( \widehat{H}=\widehat{H}_{0}+\alpha \overrightarrow{Z}\cdot (%
\overrightarrow{p}\times \overrightarrow{\sigma })\right) $\textrm{\ }and 3d
Luttinger\textrm{\ }$\left( \widehat{H}=\widehat{H}_{0}\widehat{I}%
_{4}+\lambda (\overrightarrow{p}\cdot \overrightarrow{S})^{2}\right) $%
\textrm{\ }model\textrm{\ }Hamiltonians, where\textrm{\ }$\overrightarrow{S}$%
\textrm{\ }is the\textrm{\ }$3/2$\textrm{\ }angular momentum operator and%
\textrm{\ }$\widehat{I}_{4}$\textrm{\ }is\textrm{\ }a $4\times 4$\textrm{\ }%
identity matrix. It is found that no extra term of particle current appears
for these two\textrm{\ }models\textrm{. }The main source to induce an extra
term of particle current for Dresselhaus Hamiltonian seems from the triple
power of momentum operator in its Hamiltonian. Indeed,\textrm{\ }in general,
if the Hamiltonian includes momentum operator with order higher than power $%
2 $, the extra term of particle current will appear. For Rashba and
Luttinger models, the highest power of momentum operator is $2$, but
Dresselhaus is $3$. So this is the reason why the extra term of particle
current appears in Dresselhaus, but not in Rashba and Luttinger systems. In
section 5, we will show that from an extended Noether's theorem.

For showing such argument more clearly, we study a simple $p^{4}$ model
Hamiltonian without SOC:
\begin{eqnarray}
\widehat{H} &=&\widehat{H}_{0}+\widehat{H}_{I}+V(r),  \notag \\
\widehat{H}_{I} &=&\beta \widehat{p}^{4},  \label{8}
\end{eqnarray}%
where $\beta $ is a coupling constant. In section 4, we will show that from
the extended Noether's thoerem%
\begin{eqnarray}
\frac{\partial n(\overrightarrow{r},t)}{\partial t} &=&-\nabla \cdot (%
\overrightarrow{j}_{conv}+\overrightarrow{j}_{extra}),  \notag \\
\overrightarrow{j}_{extra} &=&2\beta \hbar ^{3}\sum\limits_{n}\rho _{n}{Im}%
\{\nabla \left( \Psi _{n}^{\dag }\nabla ^{2}\Psi _{n}\right) \}  \label{9}
\end{eqnarray}%
For nonhomogeneous system, $\Psi _{n}^{\dag }\nabla ^{2}\Psi _{n}$ is
complex including both real and imaginary parts in general. $\overrightarrow{%
j}_{conv}$ and $\overrightarrow{j}_{extra}$ in equation (9) are position
dependent, $\nabla \cdot \overrightarrow{j}_{conv}$ and $\nabla \cdot
\overrightarrow{j}_{extra}$ can be nonzero in general case. Therefore, they
do not satisfy the continuity equation separately. But the combination of
them gives the following continuity equation for any stead state $(\frac{%
\partial n(\overrightarrow{r},t)}{\partial t}=0).$\ \
\begin{equation}
\nabla \cdot \overrightarrow{j}=\nabla \cdot \overrightarrow{j}%
_{conv}+\nabla \cdot \overrightarrow{j}_{extra}=0  \label{10}
\end{equation}%
In this example, we clearly show that the extra term of particle current
comes from the part of Hamiltonian with momentum $\overrightarrow{p}^{n}$ $%
(n=4).$

\section{Spin- and total angular momentum current}

When the operator $\widehat{A}$ in expression (1) is spin $\widehat{S}^{z}$
or the total angular momentum (TAM), $\widehat{L}^{z}=\widehat{l}^{z}+%
\widehat{S}^{z},$ equation (2) defines the time evolution of spin-density or
the TAM density that could relate to the spin-current density or total
angular momentum current density (TAMCD) both often appeared in recent
literatures [7-10]. For the Rashba Hamiltonian, $[\widehat{S}^{z},\widehat{H}%
_{R}]_{-}\neq 0,$ the spin $\widehat{S}^{z}$ is not a conserved quantity.
Ref.[8,9] have shown that
\begin{eqnarray}
\frac{\partial w_{\widehat{S}^{z}}^{(n)}(\overrightarrow{r},t)}{\partial t}
&=&{Re}\frac{\partial }{\partial t}[\Psi _{n}^{\dagger }\widehat{S}^{z}\Psi
_{n}]  \notag \\
&=&-\nabla \cdot {Re}\{\Psi _{n}^{\dagger }\left( \frac{1}{i\hbar }[%
\overrightarrow{r},\widehat{H}]\widehat{S}^{z}\Psi _{n}\right) \}  \notag \\
&&+{Re}\{\Psi _{n}^{\dagger }\left( (\alpha (\overrightarrow{\widehat{p}}%
\times \overrightarrow{z})\times \overrightarrow{\sigma })\Psi _{n}\right) \}
\notag \\
&=&-\nabla \cdot \overrightarrow{j}_{S}^{(n)}+\overrightarrow{j}_{\omega
}^{(n)}  \label{11}
\end{eqnarray}%
$\overrightarrow{j}_{S}=\sum\limits_{n}\rho _{n}\overrightarrow{j}_{S}^{(n)}$
is so called "spin current" widely in recent literatures,
\begin{equation*}
\overrightarrow{j}_{\omega }=\sum\limits_{n}\rho _{n}\overrightarrow{j}%
_{\omega }^{(n)}={Re}\left\{ \sum\limits_{n}\rho _{n}\Psi _{n}^{\dagger
}\left( (\alpha (\overrightarrow{\widehat{p}}\times \overrightarrow{z}%
)\times \overrightarrow{\sigma })\Psi _{n}\right) \right\}
\end{equation*}%
which is called as spin "torque". The conventional spin current\ density $%
\overrightarrow{j}_{S}$ just partly contribute to the time evolution of
spin\ density $w_{\widehat{S}^{z}}(\overrightarrow{r},t)(=\sum\limits_{n}%
\rho _{n}w_{\widehat{S}^{z}}^{(n)}(\overrightarrow{r},t)).$ Even $\partial
w_{\widehat{S}^{z}}(\overrightarrow{r},t)/\partial t=0$ for a steady state,
the conventional spin current\ density $\overrightarrow{j}_{S}$ is not
conserved ( $\nabla \cdot \overrightarrow{j}_{conv}\neq 0$ ) and satisfies
the equation $\nabla \cdot \overrightarrow{j}_{conv}=\overrightarrow{j}%
_{\omega }.$

Now we study the current of TAM $J^{z}=l^{z}+S^{z}$ in Rashba system, where $%
l^{z}$ is the $z$ component of orbital angular momentum. The case is
different from particle current. In case of particle current, the number of
particles is conserved, but relative operator of particle number is momentum
operator independent, $A=N\widehat{I}.$ Here $\widehat{J}^{z}$ is also a
conserved quantity since $[\widehat{J}^{z},\widehat{H}]_{-}=0$ for Rashba
Hamiltonian, but it contains the momentum operator. So we would check
whether its related current has extra term even for Rashba Hamiltonian
without higher than 2 power of momentum operator. After long careful
calculations (see appendix C), we obtain
\begin{equation*}
\partial _{t}\sum_{n}\rho _{n}w_{\widehat{J}^{z}}^{(n)}(\overrightarrow{r}%
,t)=\partial _{t}\sum\limits_{n}\rho _{n}{Re}[\Psi _{n}^{\dagger }\widehat{J}%
^{z}\Psi _{n}]=-\nabla \cdot \overrightarrow{j}_{L},
\end{equation*}%
where%
\begin{eqnarray}
\overrightarrow{j}_{L} &=&{Re}\sum_{n}\rho _{n}\left\{ \Psi _{n}^{\dagger }(%
\overrightarrow{r},t)\frac{1}{i\hbar }[\overrightarrow{r},\widehat{H}]%
\widehat{J}^{z}\Psi _{n}\right\}  \notag \\
&&+\frac{1}{2}\frac{\hbar ^{2}}{2m^{\ast }}\sum_{n}\rho _{n}\left\{ \Psi
_{n}^{\dagger }\left( e_{x}\nabla _{y}-e_{y}\nabla _{x}\right) \Psi
_{n}+(\left( e_{x}\nabla _{y}-e_{y}\nabla _{x}\right) \Psi _{n})^{\dag }\Psi
_{n}\right\}  \notag \\
&&+\frac{1}{2}\frac{\hbar ^{2}}{2m^{\ast }}\sum_{n}\rho _{n}\{(\nabla \Psi
_{n}^{\dagger })\left( x\nabla _{y}-y\nabla _{x}\right) \Psi _{n}+(\left(
x\nabla _{y}-y\nabla _{x}\right) \Psi _{n})^{\dag }\nabla \Psi _{n}  \notag
\\
&&+\Psi _{n}^{\dagger }\left( x\nabla _{y}-y\nabla _{x}\right) \nabla \Psi
_{n}+(\left( x\nabla _{y}-y\nabla _{x}\right) \nabla \Psi _{n})^{\dag }\Psi
_{n}\}  \label{12}
\end{eqnarray}%
The conventional definition for $\widehat{J}^{z}$ current density in
literatures is defined by%
\begin{equation*}
\overrightarrow{j}_{conv}=\frac{1}{2}\sum\limits_{n}\rho _{n}\left\{ \Psi
_{n}^{\dagger }\left[ \frac{1}{i\hbar }[\overrightarrow{r},\widehat{H}]%
\widehat{J}^{z}+\widehat{J}^{z}\frac{1}{i\hbar }[\overrightarrow{r},\widehat{%
H}]\right] \Psi _{n}\right\} .
\end{equation*}%
After taking the real part, it becomes:
\begin{equation*}
\overrightarrow{j}_{conv}=\frac{1}{2}{Re}\sum\limits_{n}\rho _{n}\left\{
\Psi _{n}^{\dagger }\left[ \frac{1}{i\hbar }[\overrightarrow{r},\widehat{H}]%
\widehat{J}^{z}+\widehat{J}^{z}\frac{1}{i\hbar }[\overrightarrow{r},\widehat{%
H}]\right] \Psi _{n}\right\} .
\end{equation*}%
Now we use the following relation%
\begin{equation}
\left[ J_{z},\frac{1}{i\hbar }[\overrightarrow{r},\widehat{H}]\right]
=i\hbar \lbrack (p_{y}/m^{\ast }-\alpha \sigma _{x})\overrightarrow{e}%
_{x}-(p_{x}/m^{\ast }+\alpha \sigma _{y})\overrightarrow{e}_{y}]  \notag
\end{equation}%
then change the form of $\overrightarrow{j}_{conv}$\ into%
\begin{eqnarray}
\overrightarrow{j}_{conv} &=&{Re}\sum\limits_{n}\rho _{n}\{\Psi
_{n}^{\dagger }\frac{1}{i\hbar }[\overrightarrow{r},\widehat{H}]\widehat{J}%
^{z}\Psi _{n}\}  \notag \\
&&+\frac{1}{2}{Re}\sum\limits_{n}\rho _{n}\left\{ i\hbar \Psi _{n}^{\dagger
}[(p_{y}/m^{\ast }-\alpha \sigma _{x})\overrightarrow{e}_{x}-(p_{x}/m^{\ast
}+\alpha \sigma _{y})\overrightarrow{e}_{y}]\Psi _{n}\right\}  \notag \\
&=&{Re}\sum\limits_{n}\rho _{n}\{\Psi _{n}^{\dagger }\frac{1}{i\hbar }[%
\overrightarrow{r},\widehat{H}]\widehat{J}^{z}\Psi _{n}\}+\frac{1}{2}{Re}%
\sum\limits_{n}\rho _{n}\{i\hbar \Psi _{n}^{\dagger }[\frac{p_{y}}{m^{\ast }}%
\overrightarrow{e}_{x}-\frac{p_{x}}{m^{\ast }}\overrightarrow{e}_{y}]\Psi
_{n}\}  \label{13}
\end{eqnarray}%
In the last step of above equation, we have used that $\Psi _{n}^{\dagger }%
\overrightarrow{\sigma }\Psi _{n}$ are real since $\{\sigma _{x},\sigma
_{y},\sigma _{z}\}$ are Hermitian and trace has been taken in spin space (
but be care for that no trace taken in position space). In appendix C, we
obtain%
\begin{equation}
\overrightarrow{j}_{extra}=\overrightarrow{j}_{L}-\overrightarrow{j}_{conv}=-%
\frac{1}{4m^{\ast }}\sum_{n}\rho _{n}l^{z}\overrightarrow{p}(\Psi
_{n}^{\dagger }\Psi _{n})\neq 0  \label{14}
\end{equation}%
Thus, we conclude that conventional formula of TAMCD, $\overrightarrow{j}%
_{conv},$ is not complete. It must have an extra term even for Rashba
system. Therefore, we suggest that one should be careful in using the
conventional current formula and do the deduction of formula carefully for
every new type Hamiltonian, particularly for some Hamiltonians with SOC in
semiconductors. In this note, the TAM, $\langle J_{L}^{Z}\rangle ,$ is not a
conserved quantity for Dresselhaus Hamiltonian since $[J_{L}^{Z},\widehat{H}%
_{Dresselhaus}]\_\neq 0.$ So, we did not attempt to check the existence of
its extra part.

\section{Noether's Theorem and Extra Term of Current}

Based on time dependent Schr\"{o}dinger equations and the particle number
conservation (see section 2), we have derived new expressions of conserved
particle current for some Hamiltonians, which include term\textsl{\ }that
the order of momentum operators in it is higher than $2$. In parallel to
that, in fact, expressions of particle current can also be derived by
Noether's theorem in field theory from $U\left( 1\right) $ gauge invariance
of Lagrangian. However, the usual Noether's theorem only consider systems
with term $p^{2}/2m$ in the momentum dependent part of Hamiltonian. So%
\textsl{\ }the Lagrangian of the system is only a functional of field
functions and their first order derivatives, $\phi (x_{\mu }),\phi ^{\dagger
}(x_{\mu }),\partial _{\mu }\phi (x_{\mu })$ and $\partial _{\mu }\phi
^{\dagger }(x_{\mu })$, for a complex scalar field, where $x_{\mu }=\left(
t,x,y,z\right) ,\mu =0,1,2,3;(\partial _{0},\partial _{1},\partial
_{2},\partial _{3})=(\partial _{t},\partial _{x},\partial _{y},\partial
_{z}).$ Simply, we denote this Lagrange as $\mathcal{L}(\phi ,\phi ^{\dagger
},\partial _{\mu }\phi ,\partial _{\mu }\phi ^{\dagger }).$ Let us briefly
recall the usual Noether theorem. An action is defined by
\begin{equation*}
S=\int dx^{4}\mathcal{L}(\phi ,\phi ^{\dagger },\partial _{\mu }\phi
,\partial _{\mu }\phi ^{\dagger }),dx^{4}=\prod_{\mu }dx_{\mu }.
\end{equation*}%
\begin{eqnarray}
\delta S &=&\int dx^{4}\left\{ \left( \frac{\partial \mathcal{L}}{\partial
\phi }\delta \phi +\sum_{\mu }\frac{\partial \mathcal{L}}{\partial (\partial
_{\mu }\phi )}\delta \partial _{\mu }\phi \right) +\left( \phi \rightarrow
\phi ^{\dagger }\right) \right\}  \notag \\
&=&\int dx^{4}\left\{ \left( \frac{\partial \mathcal{L}}{\partial \phi }%
-\sum_{\mu }\partial _{\mu }\frac{\partial \mathcal{L}}{\partial (\partial
_{\mu }\phi )}\right) \delta \phi +\left( \phi \rightarrow \phi ^{\dagger
}\right) \right\}  \notag \\
&&+\int dx^{4}\left\{ \sum_{\mu }\partial _{\mu }\left( \frac{\partial
\mathcal{L}}{\partial (\partial _{\mu }\phi )}\delta \phi \right) +\left(
\phi \rightarrow \phi ^{\dagger }\right) \right\} .  \label{15}
\end{eqnarray}%
Where two variational functions $\delta \phi $ and $\delta \phi ^{\dagger }$
are independent. According to the least action principle, $\delta S=0,$
which gives two conjugate dynamical equations of $\phi $ and $\phi ^{\dagger
}$. One is from the term of $\delta \phi $ and another from $\delta \phi
^{\dagger }.$ For a complex scalar field, the Lagrangian is usually given by%
\begin{equation*}
\mathcal{L(}\phi ,\phi ^{\dagger },\partial _{\mu }\phi ,\partial _{\mu
}\phi ^{\dagger }\mathcal{)}=\phi ^{\dagger }i\partial _{0}\phi -\frac{1}{2m}%
\sum_{i=1,2,3}\left( \partial _{i}\phi \right) ^{\dagger }\left( \partial
_{i}\phi \right) -\phi ^{\dagger }V(\{x_{\mu }\})\phi ,
\end{equation*}%
where $\hbar =1.$ From equation (15), we have%
\begin{eqnarray*}
\frac{\partial \mathcal{L}}{\partial \phi ^{\dagger }}-\left( \partial _{\mu
}\frac{\partial \mathcal{L}}{\partial \left( \partial _{\mu }\phi ^{\dagger
}\right) }\right) &=&0\rightarrow \\
i\partial _{0}\phi +\frac{1}{2m}\partial _{i}^{2}\phi -V(\{x_{\mu }\})\phi
&=&0.
\end{eqnarray*}%
It is Schr\"{o}dinger equation:%
\begin{eqnarray*}
i\frac{\partial }{\partial t}\phi &=&H\phi , \\
H &=&-\frac{\nabla ^{2}}{2m}+V(\{x_{\mu }\}),
\end{eqnarray*}%
Equation (15) also yields an equation:
\begin{equation}
\sum\limits_{\mu =0,1,2,3}\partial _{\mu }F_{\mu }=0.  \label{16}
\end{equation}%
which defines a $4D$ current:%
\begin{equation}
F_{\mu }=\left( \frac{\partial \mathcal{L}}{\partial (\partial _{\mu }\phi )}%
\delta \phi \right) +\left( \phi \rightarrow \phi ^{\dagger }\right)
\label{17}
\end{equation}%
Input usual Lagrangian $\mathcal{L}$ into equation (17), we have%
\begin{eqnarray*}
F_{0} &=&i\phi ^{\dagger }\delta \phi , \\
F_{i} &=&-\frac{1}{2m}[(\partial _{i}\phi )^{\dagger }\delta \phi +(\partial
_{i}\phi )\delta \phi ^{\dagger }],i=1,2,3.
\end{eqnarray*}%
By $U(1)$ gauge transformation
\begin{eqnarray*}
\phi &\rightarrow &e^{i\alpha }\phi \approx (1+i\alpha )\phi =\phi +i\alpha
\phi \rightarrow \\
\delta \phi &=&i\alpha \phi ,\delta \phi ^{\dagger }=-i\alpha \phi ^{\dagger
},
\end{eqnarray*}%
and due to $U(1)$ gauge invariance of the Hamiltonian, we obtain
\begin{eqnarray*}
F_{0} &=&-\alpha \phi ^{\dagger }\phi \\
F_{i} &=&-\frac{i\alpha }{2m}\left\{ (\partial _{i}\phi )^{\dagger }\phi
-\phi ^{\dagger }\partial _{i}\phi \right\} =\frac{\alpha }{m}{Re}\phi
^{\dagger }i\partial _{i}\phi \\
&=&-\alpha {Re}\left\{ \phi ^{\dagger }\frac{1}{i}[r_{i},H]_{-}\phi \right\}
=-\alpha {Re}\left\{ \phi ^{\dagger }v_{i}\phi \right\} ,i=1,2,3.
\end{eqnarray*}%
We obtain the continuity equation of particle%
\begin{equation}
\frac{\partial \left( \phi ^{\dagger }\phi \right) }{\partial t}+\widehat{%
\nabla }\cdot {Re}\left\{ \phi ^{\dagger }\frac{1}{i}[\overrightarrow{r}%
,H]_{-}\phi \right\} =0.  \label{18}
\end{equation}%
This is just the conventional expression of particle current.
\begin{equation}
\overrightarrow{j}_{conv}={Re}\left\{ \phi ^{\dagger }\frac{1}{i}[%
\overrightarrow{r},H]_{-}\phi \right\}  \label{19}
\end{equation}%
For Rashba and Luttinger systems, the usual Noether's theorem can be applied
since the additional spin-orbit-coupling term in these two Hamiltonians are
momentum operator $p$ or $p^{2}$ dependent. Therefore, it can be unstood
that why their expressions of the particle currents are also conventional.
It is consistent with our conclusion. However, when we study the system like
$H=p^{2}/2m+\beta p^{4}$ or $H_{Dresselhaus}$,\textsl{\ }the terms of high
order derivatives, inherited from Hamiltonian, should also be included in
the Lagrangian:\textsl{\ }$\mathcal{L}(\phi ,\partial _{\mu }\phi ,\partial
_{\mu }\partial _{\nu }\phi ,\cdot \cdot \cdot ;\left( \phi \rightarrow \phi
^{\dagger }\right) )$, thus the usual Noether's theorem should be extended.
According to variational method,
\begin{eqnarray*}
\delta S &=&\int_{V_{4}}dx^{4}\delta \mathcal{L}([\phi ,\partial _{\mu }\phi
,\partial _{\mu }\partial _{\nu }\phi ,\cdot \cdot \cdot ;\left( \phi
\rightarrow \phi ^{\dagger }\right) ]) \\
&=&\int_{V_{4}}dx^{4}\left\{ \frac{\partial \mathcal{L}}{\partial \phi }%
\delta \phi +\frac{\partial \mathcal{L}}{\partial (\partial _{\mu }\phi )}%
\delta (\partial _{\mu }\phi )+\frac{\partial \mathcal{L}}{\partial
(\partial _{\mu }\partial _{\nu }\phi )}\delta (\partial _{\mu }\partial
_{\nu }\phi )+\cdot \cdot \cdot +\left( \phi \rightarrow \phi ^{\dagger
}\right) \right\} \\
&=&\int_{V_{4}}dx^{4}\left\{ \frac{\partial \mathcal{L}}{\partial \phi }%
\delta \phi -\partial _{\mu }\left( \frac{\partial \mathcal{L}}{\partial
(\partial _{\mu }\phi )}\right) \delta \phi +\partial _{\nu }\partial _{\mu
}\left( \frac{\partial \mathcal{L}}{\partial (\partial _{\mu }\partial _{\nu
}\phi )}\right) \delta \phi +\cdot \cdot \cdot +\left( \phi \rightarrow \phi
^{\dagger }\right) \right\} \\
&&+\int_{V_{4}}dx^{4}\partial _{\mu }\left\{ \left( \frac{\partial \mathcal{L%
}}{\partial (\partial _{\mu }\phi )}\delta \phi \right) -\partial _{\nu
}\left( \frac{\partial \mathcal{L}}{\partial (\partial _{\mu }\partial _{\nu
}\phi )}\right) \delta \phi +\left( \frac{\partial \mathcal{L}}{\partial
(\partial _{\mu }\partial _{\nu }\phi )}\delta (\partial _{\nu }\phi
)\right) +\cdot \cdot \cdot +\left( \phi \rightarrow \phi ^{\dagger }\right)
\right\}
\end{eqnarray*}%
where $dx^{4}=\prod\limits_{\mu }dx_{\mu }.$ Now the $4D$ current is defined
by%
\begin{eqnarray}
F_{\mu } &=&\left( \frac{\partial \mathcal{L}}{\partial (\partial _{\mu
}\phi )}\delta \phi \right) -\partial _{\nu }\left( \frac{\partial \mathcal{L%
}}{\partial (\partial _{\mu }\partial _{\nu }\phi )}\right) \delta \phi
+\left( \frac{\partial \mathcal{L}}{\partial (\partial _{\mu }\partial _{\nu
}\phi )}\delta (\partial _{\nu }\phi )\right)  \notag \\
&&+\cdot \cdot \cdot +\left( \phi \rightarrow \phi ^{\dagger }\right)
\label{20}
\end{eqnarray}%
and
\begin{eqnarray*}
\triangle \mathcal{L} &=&\frac{\partial \mathcal{L}}{\partial \phi }%
-\partial _{\mu }\left( \frac{\partial \mathcal{L}}{\partial (\partial _{\mu
}\phi )}\right) +\partial _{\nu }\partial _{\mu }\left( \frac{\partial
\mathcal{L}}{\partial (\partial _{\mu }\partial _{\nu }\phi )}\right) \\
&&+\cdot \cdot \cdot +\left( \phi \rightarrow \phi ^{\dagger }\right)
\end{eqnarray*}%
Thus%
\begin{equation*}
\delta S=\int_{V_{4}}dx^{4}\triangle \mathcal{L}\delta \phi
+\int_{V_{4}}dx^{4}\partial _{\mu }F_{\mu }=0.
\end{equation*}%
According to the least action principle, a suitable definition of Lagrangian
is necessary for obtaining the correct dynamical equation of $\phi $ from $%
\triangle \mathcal{L}=0$ and conservation law from $\sum_{\mu }\partial
_{\mu }F_{\mu }=0.$ Therefore, the expression of particle current $%
\overrightarrow{j}$ can be obtained from the invariance of $U(1)$ gauge
transformation for such extended system:%
\begin{eqnarray}
j_{i} &=&\left( \frac{\partial \mathcal{L}}{\partial (\partial _{i}\phi )}%
\delta \phi \right) -\partial _{\nu }\left( \frac{\partial \mathcal{L}}{%
\partial (\partial _{i}\partial _{\nu }\phi )}\right) \delta \phi +\left(
\frac{\partial \mathcal{L}}{\partial (\partial _{i}\partial _{\nu }\phi )}%
\delta (\partial _{\nu }\phi )\right)  \notag \\
&&+\cdot \cdot \cdot +\left( \phi \rightarrow \phi ^{\dagger }\right)
\label{21}
\end{eqnarray}%
Comparing equation (21) with (15), there are some extra terms $F_{\mu
}^{extra}:$%
\begin{eqnarray*}
F_{\mu }^{extra} &=&-\partial _{\nu }\left( \frac{\partial \mathcal{L}}{%
\partial (\partial _{\mu }\partial _{\nu }\phi )}\right) \delta \phi +\left(
\frac{\partial \mathcal{L}}{\partial (\partial _{\mu }\partial _{\nu }\phi )}%
\delta (\partial _{\nu }\phi )\right) \\
&&+\cdot \cdot \cdot +\left( \phi \rightarrow \phi ^{\dagger }\right)
\end{eqnarray*}%
This is the reason why extra terms do exist in particle current for the $%
p^{4}$ model and $p^{3}$-Dresselhaus Hamiltonian. In this paper, as an
example, we use the above extended Noether's theorem to study the $p^{4}$
model system in detail only. Further study on Noether theorem for other
systems must be interesting and it is ongoing.

\ The central point of Noether's theorem is to find out a suitable
functional expression of Lagrangian, then one can derive the time dependent
Schr\"{o}dinger equation and conservation law. For $p^{4}$ mode, we choose
the following symmetric Lagrangian%
\begin{eqnarray}
&&\mathcal{L}[\phi \left( x\right) ,\partial _{\mu }\phi \left( x\right)
,\partial _{\mu }^{2}\phi \left( x\right) ,\partial _{\nu }\partial _{\mu
}^{2}\phi \left( x\right) ,\partial _{\nu }^{2}\partial _{\mu }^{2}\phi
\left( x\right) ;\left( \phi \rightarrow \phi ^{\dagger }\right) ]  \notag \\
&=&\phi ^{\dagger }\left( i\partial _{0}\phi \right) -\frac{1}{2m}\left(
\partial _{i}\phi \right) ^{\dagger }\left( \partial _{i}\phi \right) -\frac{%
1}{5}\beta \cdot  \notag \\
&&[\left( \partial _{i}^{2}\partial _{j}^{2}\phi \right) ^{\dagger }\phi
-\left( \partial _{i}^{2}\partial _{j}\phi \right) ^{\dagger }\left(
\partial _{j}\phi \right) +\left( \partial _{i}^{2}\phi \right) ^{\dagger
}\left( \partial _{j}^{2}\phi \right)  \notag \\
&&-\left( \partial _{i}\phi \right) ^{\dagger }\left( \partial _{i}\partial
_{j}^{2}\phi \right) +\phi ^{\dagger }\left( \partial _{i}^{2}\partial
_{j}^{2}\phi \right) ],  \label{22}
\end{eqnarray}%
After long algebra, but no difficulty, we obtain Schr\"{o}dinger equation%
\begin{equation*}
i\frac{\partial }{\partial t}\phi =\left( \frac{p^{2}}{2m}+\beta
p^{4}\right) \phi ,
\end{equation*}%
and the following equation for $4D$ conserved current%
\begin{eqnarray}
F_{\mu } &=&[\frac{\partial \mathcal{L}}{\partial \left( \partial _{\mu
}\phi \right) }\delta \phi +\frac{\partial \mathcal{L}}{\partial \left(
\partial _{\mu }^{2}\phi \right) }\delta \left( \partial _{\mu }\phi \right)
-(\partial _{\mu }\frac{\partial \mathcal{L}}{\partial \left( \partial _{\mu
}^{2}\phi \right) })\delta \phi  \notag \\
&&+\frac{\partial \mathcal{L}}{\partial \left( \partial _{\nu }\partial
_{\mu }^{2}\phi \right) }\delta \left( \partial _{\nu }\partial _{\mu }\phi
\right) -(\partial _{\mu }\frac{\partial \mathcal{L}}{\partial \left(
\partial _{\nu }\partial _{\mu }^{2}\phi \right) })\delta \left( \partial
_{\nu }\phi \right)  \notag \\
&&+(\partial _{\nu }^{2}\frac{\partial \mathcal{L}}{\partial \left( \partial
_{\mu }\partial _{\nu }^{2}\phi \right) }\delta \phi +\frac{\partial
\mathcal{L}}{\partial \left( \partial _{\nu }^{2}\partial _{\mu }^{2}\phi
\right) }\delta \left( \partial _{\nu }^{2}\partial _{\mu }\phi \right)
\notag \\
&&-(\partial _{\mu }\frac{\partial \mathcal{L}}{\partial \left( \partial
_{\nu }^{2}\partial _{\mu }^{2}\phi \right) })\delta \left( \partial _{\nu
}^{2}\phi \right) +(\partial _{\nu }^{2}\frac{\partial \mathcal{L}}{\partial
\left( \partial _{\mu }^{2}\partial _{\nu }^{2}\phi \right) })\delta \left(
\partial _{\mu }\phi \right)  \notag \\
&&-(\partial _{\nu }^{2}\partial _{\mu }\frac{\partial \mathcal{L}}{\partial
\left( \partial _{\mu }^{2}\partial _{\nu }^{2}\phi \right) })\delta \phi
]+\left( \phi \rightarrow \phi ^{\dagger }\right) ,  \label{23}
\end{eqnarray}%
and conservation law $\partial _{\mu }F_{\mu }=0.$ It is an extension of the
conventional expression $F_{\mu }=\frac{\partial \mathcal{L}}{\partial
\left( \partial _{\mu }\phi \right) }\delta \phi +h.c.$ for usual
Hamiltonian. Now applying $U\left( 1\right) $ gauge symmetry of the
Lagrangian to eq.(22), performing transformation: $\phi \longrightarrow $ $%
\phi +i\alpha \phi ,\phi ^{\dagger }\longrightarrow $ $\phi ^{\dagger
}-i\alpha \phi ^{\dagger },$ where $\alpha $ is a small constant, we gain
the expression of conserved particle current%
\begin{eqnarray}
\overrightarrow{j}_{cons} &=&\frac{-i}{2m}[\phi ^{\dagger }\left( \nabla
\phi \right) -\left( \nabla \phi \right) ^{\dagger }\phi ]+i2\beta \lbrack
\phi ^{\dagger }\left( \nabla \nabla ^{2}\phi \right) -\left( \nabla \nabla
^{2}\phi ^{\dagger }\right) \phi ]  \notag \\
&&-i\left( \frac{1}{5}\beta \right) [-5\left( \nabla \nabla ^{2}\phi
^{\dagger }\right) \phi +5\phi ^{\dagger }\left( \nabla \nabla ^{2}\phi
\right)  \notag \\
&&+3\left( \nabla \phi ^{\dagger }\right) \left( \nabla ^{2}\phi \right)
+2\left( \nabla \phi \right) ^{\dagger }\cdot \left( \nabla \nabla \phi
\right)  \notag \\
&&-3\left( \nabla ^{2}\phi \right) ^{\dagger }\left( \nabla \phi \right)
-2\left( \nabla \nabla \phi ^{\dagger }\right) \cdot \left( \nabla \phi
\right) ]  \label{24}
\end{eqnarray}%
which is essentially the same as eq.$\left( 9\right) $ in Section 2. We have
proved that $\nabla \cdot \left( \overrightarrow{j}_{cons}-(\overrightarrow{j%
}_{conv}+\overrightarrow{j}_{extra})\right) =0$.

We also have proved that a suitable symmetric Lagrangian for $p^{3}$
Dresselhaus model can be found and correct Schr\"{o}dinger equation and
particle current are gained. The expression of particle current is
essentially the same as the one in section 2 with a difference $\triangle
\overrightarrow{j}_{Dresselhaus}=\nabla \times \overrightarrow{A},$ a $rot$
vector, only, so $\nabla \cdot \left( \triangle \overrightarrow{j}%
_{Dresselhaus}\right) =0$. Further studies on the extension of Noether's
theorem about the construction of Lagrangian for such unusual system, the
multiple correspondence from Hamiltonian to Lagrangian and finding general
approach is our next studies.\bigskip

In summary: we have studied the conventional formula of particle\ density
for some Hamiltonians such as Rashba, Luttinger, Dresselhaus Hamiltonian as
well as a $p^{4}$ model Hamiltonian. It is shown that the conventional
formula of particle current\ density is correct for Rashba and Luttinger
Hamiltonians. But it lacks an additional term of current\ density $%
\overrightarrow{j}_{extra}$ for $p^{3}$ Dresselhaus Hamiltonian and $p^{4}$
one without SOC. The spin and TAM currents are also addressed for Rashba
Hamiltonian. Some detailed proofs are presented in appendices to show the
existence of extra terms of currents. The results point out that the
conventional expression of TAMCD is not complete even for Rashba term. It
against the conservation law of the total angular momentum if the extra term
is neglected. In deed, the extra term saves the conservation of total
angular momentum for Rashba system. In the final, a brief explanation of the
extension of Noether's theorem is presented. It just shows that the
deviation of the expression of particle current from conventional formula
derived by Noether's theorem for some system with SOC is reasonable and no
doubt. The general extension of Noether's theorem is still not quite clear
and in studying. However, our approach to derive the conserved currents is
based on Hamiltonian and time dependent Schr\"{o}dinger equation, our
results are rigorous and unique up to a $rot$ vector (divergence =0) without
any doubt. A correct formula of particle current is important in studying
transport problems in recent SOC systems.

\textit{Acknowledgement}: The work is supported by National Natural Science
Foundation of China (No.10674027) and 973 Project of China (No.
2002CB613504).

\section{Appendix}

\subsection{Extra term of particle current for Dresselhaus system}

In this part, we would show some details leading to the extra term in the
current density for Dresselhaus Hamiltonian $\widehat{H}=\widehat{H}_{0}+%
\widehat{H}_{D}$. Let's begin from the equation of (6) in the body of the
text
\begin{align}
\frac{\partial n}{\partial t}& =-(\frac{1}{2}\frac{1}{i\hbar }\nabla \cdot
\lbrack \overrightarrow{r},\widehat{H}_{0}]\Psi _{n})^{\dagger }\Psi
_{n}-\Psi _{n}^{\dag }(\frac{1}{2}\frac{1}{i\hbar }\nabla \cdot \lbrack
\overrightarrow{r},\widehat{H}_{0}])\Psi _{n})  \notag \\
& +\left( \frac{1}{i\hbar }\widehat{H}_{D}\Psi _{n}(\overrightarrow{r}%
,t)\right) ^{\dagger }\Psi _{n}(\overrightarrow{r},t)+\Psi _{n}^{\dag
}\left( \frac{1}{i\hbar }\widehat{H}_{D}\Psi _{n}(\overrightarrow{r}%
,t)\right) .  \tag{A.1}
\end{align}%
Now we focus on the deduction of the last two terms of Eq. (A.1).
Considering SOC part of Hamiltonian%
\begin{equation*}
\widehat{H}_{D}=\eta \left[ p_{x}\left( p_{y}^{2}-p_{z}^{2}\right) \sigma
_{x}+p_{y}\left( p_{z}^{2}-p_{x}^{2}\right) \sigma _{y}+p_{z}\left(
p_{x}^{2}-p_{y}^{2}\right) \sigma _{z}\right] ,
\end{equation*}%
we obtain%
\begin{equation}
\left( \frac{1}{i\hbar }\left[ x,\widehat{H}_{D}\right] +2\eta \left(
p_{y}p_{x}\sigma _{y}-p_{z}p_{x}\sigma _{z}\right) \right) =\eta \left(
p_{y}^{2}-p_{z}^{2}\right) \sigma _{x},  \tag{A.2}
\end{equation}%
\begin{equation}
\left( \frac{1}{i\hbar }\left[ y,\widehat{H}_{D}\right] +2\eta \left(
p_{z}p_{y}\sigma _{z}-p_{x}p_{y}\sigma _{x}\right) \right) =\eta \left(
p_{z}^{2}-p_{x}^{2}\right) \sigma _{y},  \tag{A.3}
\end{equation}%
\begin{equation}
\left( \frac{1}{i\hbar }\left[ z,\widehat{H}_{D}\right] +2\eta \left(
p_{x}p_{z}\sigma _{x}-p_{y}p_{z}\sigma _{y}\right) \right) =\eta \left(
p_{x}^{2}-p_{y}^{2}\right) \sigma _{z}.  \tag{A.4}
\end{equation}%
\begin{eqnarray*}
\widehat{H}_{D} &=&\overrightarrow{p}\cdot \frac{1}{i\hbar }\left[
\overrightarrow{r},\widehat{H}_{D}\right] +2\eta \left( p_{y}\sigma
_{y}-p_{z}\sigma _{z}\right) p_{x}^{2} \\
&&+2\eta \left( p_{z}\sigma _{z}-p_{x}\sigma _{x}\right) p_{y}^{2}+2\eta
\left( p_{x}\sigma _{x}-p_{y}\sigma _{y}\right) p_{z}^{2} \\
&=&\overrightarrow{p}\cdot \frac{1}{i\hbar }\left[ \overrightarrow{r},%
\widehat{H}_{D}\right] -2\widehat{H}_{D}. \\
\widehat{H}_{D} &=&-\frac{1}{3}\nabla \cdot \left[ \overrightarrow{r},%
\widehat{H}_{D}\right]
\end{eqnarray*}%
Then we have
\begin{eqnarray*}
&&\left( \frac{1}{i\hbar }\widehat{H}_{D}\Psi _{n}\right) ^{\dagger }\Psi
_{n}+\Psi _{n}^{\dag }\left( \frac{1}{i\hbar }\widehat{H}_{D}\Psi _{n}\right)
\\
&=&-\frac{1}{3}\left\{ \left( \frac{1}{i\hbar }\nabla \cdot \left[
\overrightarrow{r},\widehat{H}_{D}\right] \Psi _{n}\right) ^{\dagger }\Psi
_{n}+\Psi _{n}^{\dag }\left( \frac{1}{i\hbar }\nabla \cdot \left[
\overrightarrow{r},\widehat{H}_{D}\right] \Psi _{n}\right) \right\} \\
&=&-\frac{1}{3}\nabla \cdot \left\{ \left( \frac{1}{i\hbar }\left[
\overrightarrow{r},\widehat{H}_{D}\right] \Psi _{n}\right) ^{\dagger }\Psi
_{n}+\Psi _{n}^{\dag }\left( \frac{1}{i\hbar }\left[ \overrightarrow{r},%
\widehat{H}_{D}\right] \Psi _{n}\right) \right\} \\
&&+\frac{1}{3}\left\{ \left( \frac{1}{i\hbar }\left[ \overrightarrow{r},%
\widehat{H}_{D}\right] \Psi _{n}\right) ^{\dagger }\cdot \left( \nabla \Psi
_{n}\right) +\left( \nabla \Psi _{n}\right) ^{\dagger }\cdot \left( \frac{1}{%
i\hbar }\left[ \overrightarrow{r},\widehat{H}_{D}\right] \Psi _{n}\right)
\right\}
\end{eqnarray*}%
Inserting above result into the equation (A.1), we obtain%
\begin{align*}
\frac{\partial n}{\partial t}& =-\nabla \cdot {Re}\left\{ \Psi _{n}^{\dagger
}\frac{1}{i\hbar }[\overrightarrow{r},\widehat{H}]\Psi _{n}\right\} +\frac{1%
}{3}\nabla \cdot {Re}\left\{ \Psi _{n}^{\dag }\left( \frac{1}{i\hbar }\left[
\overrightarrow{r},\widehat{H}_{D}\right] \Psi _{n}\right) \right\} \\
& +\frac{2}{3}{Re}\left\{ \left( \nabla \Psi _{n}\right) ^{\dagger }\cdot
\left( \frac{1}{i\hbar }\left[ \overrightarrow{r},\widehat{H}_{D}\right]
\Psi _{n}\right) \right\}
\end{align*}%
Now we try to express $\frac{2}{3}{Re}\left\{ \left( \nabla \Psi _{n}\right)
^{\dagger }\cdot \left( \frac{1}{i\hbar }\left[ \overrightarrow{r},\widehat{H%
}_{D}\right] \Psi _{n}\right) \right\} $ by $\nabla \cdot j_{add},$ then the
analytic formula of extra contribution to particle current can be obtained.
By using (A.2), (A.2) and (A.4), $\frac{1}{i\hbar }[\overrightarrow{r},%
\widehat{H}_{D}]$ can be substituted as following
\begin{eqnarray*}
&&\frac{1}{3}\left\{ \left( \frac{1}{i\hbar }[\overrightarrow{r},\widehat{H}%
_{D}]\Psi _{n}\right) ^{\dag }\cdot \mathbf{\nabla }\Psi _{n}+\mathbf{\nabla
}\Psi _{n}^{\dag }\cdot \left( \frac{1}{i\hbar }[\overrightarrow{r},\widehat{%
H}_{D}]\Psi _{n}\right) \right\} \\
&=&-\frac{1}{3}\hbar ^{2}\eta \{\left( \left[ \left( \nabla _{y}^{2}-\nabla
_{z}^{2}\right) \sigma _{x}-2\nabla _{y}\nabla _{x}\sigma _{y}+2\nabla
_{z}\nabla _{x}\sigma _{z}\right] \Psi _{n}\right) ^{\dag }\nabla _{x}\Psi
_{n}\} \\
&&-\frac{1}{3}\hbar ^{2}\eta \{\nabla _{x}\Psi _{n}^{\dag }\left[ \left(
\nabla _{y}^{2}-\nabla _{z}^{2}\right) \sigma _{x}-2\nabla _{y}\nabla
_{x}\sigma _{y}+2\nabla _{z}\nabla _{x}\sigma _{z}\right] \Psi _{n}\} \\
&&-\frac{1}{3}\hbar ^{2}\eta \{\left( \left[ \left( \nabla _{z}^{2}-\nabla
_{x}^{2}\right) \sigma _{y}-2\nabla _{z}\nabla _{y}\sigma _{z}+2\nabla
_{x}\nabla _{y}\sigma _{x}\right] \Psi _{n}\right) ^{\dag }\nabla _{y}\Psi
_{n}\} \\
&&-\frac{1}{3}\hbar ^{2}\eta \{\nabla _{y}\Psi _{n}^{\dag }\left[ \left(
\nabla _{z}^{2}-\nabla _{x}^{2}\right) \sigma _{y}-2\nabla _{z}\nabla
_{y}\sigma _{z}+2\nabla _{x}\nabla _{y}\sigma _{x}\right] \Psi _{n}\} \\
&&-\frac{1}{3}\hbar ^{2}\eta \{\left( \left[ \left( \nabla _{x}^{2}-\nabla
_{y}^{2}\right) \sigma _{z}-2\nabla _{x}\nabla _{z}\sigma _{x}+2\nabla
_{y}\nabla _{z}\sigma _{y}\right] \Psi _{n}\right) ^{\dag }\nabla _{z}\Psi
_{n}\} \\
&&-\frac{1}{3}\hbar ^{2}\eta \{\nabla _{z}\Psi _{n}^{\dag }\left[ \left(
\nabla _{x}^{2}-\nabla _{y}^{2}\right) \sigma _{z}-2\nabla _{x}\nabla
_{z}\sigma _{x}+2\nabla _{y}\nabla _{z}\sigma _{y}\right] \Psi _{n}\}
\end{eqnarray*}%
It can be changed to
\begin{eqnarray*}
&&\frac{1}{3}\left\{ \left( \frac{1}{i\hbar }[\overrightarrow{r},\widehat{H}%
_{D}]\Psi _{n}\right) ^{\dag }\cdot \mathbf{\nabla }\Psi _{n}+\mathbf{\nabla
}\Psi _{n}^{\dag }\cdot \left( \frac{1}{i\hbar }[\overrightarrow{r},\widehat{%
H}_{D}]\Psi _{n}\right) \right\} \\
=-\frac{1}{3}\hbar ^{2}\eta &&\{\left( \left( \nabla _{y}^{2}-\nabla
_{z}^{2}\right) \sigma _{x}\Psi _{n}\right) ^{\dag }\nabla _{x}\Psi
_{n}+(\nabla _{x}\Psi _{n}^{\dag })\left( \nabla _{y}^{2}-\nabla
_{z}^{2}\right) \sigma _{x}\Psi _{n} \\
&&+\left( 2\nabla _{x}\nabla _{y}\sigma _{x}\Psi _{n}\right) ^{\dag }\nabla
_{y}\Psi _{n}+(\nabla _{y}\Psi _{n}^{\dag })2\nabla _{x}\nabla _{y}\sigma
_{x}\Psi _{n} \\
&&-\left( 2\nabla _{x}\nabla _{z}\sigma _{x}\Psi _{n}\right) ^{\dag }\nabla
_{z}\Psi _{n}-(\nabla _{z}\Psi _{n}^{\dag })2\nabla _{x}\nabla _{z}\sigma
_{x}\Psi _{n}\}+C.P.
\end{eqnarray*}%
Where $C.P.$ is cyclic permutation of index $\{x,y,z\}$. Above equation can
be changed to%
\begin{eqnarray*}
&&\frac{1}{3}\left\{ \left( \frac{1}{i\hbar }[\overrightarrow{r},\widehat{H}%
_{D}]\Psi _{n}\right) ^{\dag }\cdot \mathbf{\nabla }\Psi _{n}+\mathbf{\nabla
}\Psi _{n}^{\dag }\cdot \left( \frac{1}{i\hbar }[\overrightarrow{r},\widehat{%
H}_{D}]\Psi _{n}\right) \right\} \\
=-\frac{1}{3}\hbar ^{2}\eta &&\{\nabla _{y}[\left( \nabla _{y}\sigma
_{x}\Psi _{n}\right) ^{\dag }\nabla _{x}\Psi _{n}+\left( \nabla _{x}\Psi
_{n}^{\dag }\right) \left( \nabla _{y}\sigma _{x}\Psi _{n}\right) ] \\
&&+\nabla _{x}[\left( \nabla _{y}\sigma _{x}\Psi _{n}\right) ^{\dag }\nabla
_{y}\Psi _{n}] \\
&&-\nabla _{z}[\left( \nabla _{z}\sigma _{x}\Psi _{n}\right) ^{\dag }\nabla
_{x}\Psi _{n}+\left( \nabla _{x}\Psi _{n}^{\dag }\right) \left( \nabla
_{z}\sigma _{x}\Psi _{n}\right) ] \\
&&-\nabla _{x}[\left( \nabla _{z}\sigma _{x}\Psi _{n}\right) ^{\dag }\nabla
_{z}\Psi _{n}]\}+C.P.
\end{eqnarray*}

\begin{eqnarray*}
&=&-\frac{1}{3}\hbar ^{2}\eta \{\nabla _{x}[\left( \nabla _{y}\sigma
_{x}\Psi _{n}\right) ^{\dag }\nabla _{y}\Psi _{n}]-\nabla _{x}[\left( \nabla
_{z}\sigma _{x}\Psi _{n}\right) ^{\dag }\nabla _{z}\Psi _{n}] \\
&&-\nabla _{x}[\left( \nabla _{x}\sigma _{y}\Psi _{n}\right) ^{\dag }\nabla
_{y}\Psi _{n}+\left( \nabla _{y}\Psi _{n}^{\dag }\right) \left( \nabla
_{x}\sigma _{y}\Psi _{n}\right) ] \\
&&+\nabla _{x}[\left( \nabla _{x}\sigma _{z}\Psi _{n}\right) ^{\dag }\nabla
_{z}\Psi _{n}+\left( \nabla _{z}\Psi _{n}^{\dag }\right) \left( \nabla
_{x}\sigma _{z}\Psi _{n}\right) ]\}+C.P. \\
&=&-\nabla \cdot \overrightarrow{K}^{\left( n\right) }
\end{eqnarray*}%
where%
\begin{eqnarray*}
K_{x}^{\left( n\right) } &=&\frac{1}{3}\hbar ^{2}\eta \{\left( \nabla
_{y}\sigma _{x}\Psi _{n}\right) ^{\dag }\nabla _{y}\Psi _{n}+\left( \nabla
_{x}\sigma _{z}\Psi _{n}\right) ^{\dag }\nabla _{z}\Psi _{n}+\left( \nabla
_{z}\Psi _{n}^{\dag }\right) \left( \nabla _{x}\sigma _{z}\Psi _{n}\right) \\
&&-\left( \nabla _{z}\sigma _{x}\Psi _{n}\right) ^{\dag }\nabla _{z}\Psi
_{n}-\left( \nabla _{x}\sigma _{y}\Psi _{n}\right) ^{\dag }\nabla _{y}\Psi
_{n}-\left( \nabla _{y}\Psi _{n}^{\dag }\right) \left( \nabla _{x}\sigma
_{y}\Psi _{n}\right) \} \\
K_{y}^{\left( n\right) } &=&\frac{1}{3}\hbar ^{2}\eta \{\left( \nabla
_{z}\sigma _{y}\Psi _{n}\right) ^{\dag }\left( \nabla _{z}\Psi _{n}\right)
+\left( \nabla _{y}\sigma _{x}\Psi _{n}\right) ^{\dag }\nabla _{x}\Psi
_{n}+\left( \nabla _{x}\Psi _{n}^{\dag }\right) \left( \nabla _{y}\sigma
_{x}\Psi _{n}\right) \\
&&-\left( \nabla _{x}\sigma _{y}\Psi _{n}\right) ^{\dag }\nabla _{x}\Psi
_{n}-\left( \nabla _{y}\sigma _{z}\Psi _{n}\right) ^{\dag }\nabla _{z}\Psi
_{n}-\nabla _{z}\Psi _{n}^{\dag }\left( \nabla _{y}\sigma _{z}\Psi
_{n}\right) \} \\
K_{z}^{\left( n\right) } &=&\frac{1}{3}\hbar ^{2}\eta \{\left( \nabla
_{x}\sigma _{z}\Psi _{n}\right) ^{\dag }\left( \nabla _{x}\Psi _{n}\right)
+\left( \nabla _{z}\sigma _{y}\Psi _{n}\right) ^{\dag }\nabla _{y}\Psi
_{n}+\left( \nabla _{y}\Psi _{n}^{\dag }\right) \left( \nabla _{z}\sigma
_{y}\Psi _{n}\right) \\
&&-\left( \nabla _{y}\sigma _{z}\Psi _{n}\right) ^{\dag }\left( \nabla
_{y}\Psi _{n}\right) -\left( \nabla _{z}\sigma _{x}\Psi _{n}\right) ^{\dag
}\nabla _{x}\Psi _{n}-\left( \nabla _{x}\Psi _{n}^{\dag }\right) \left(
\nabla _{z}\sigma _{x}\Psi _{n}\right) \}
\end{eqnarray*}%
So%
\begin{equation}
\overrightarrow{j}^{\left( n\right) }=\overrightarrow{j}_{conv}^{\left(
n\right) }+\overrightarrow{j}^{\prime \left( n\right) }+\overrightarrow{K}%
^{\left( n\right) }=\overrightarrow{j}_{conv}^{\left( n\right) }+%
\overrightarrow{j}_{extra}^{\left( n\right) }  \tag{A.5}
\end{equation}%
where%
\begin{eqnarray*}
\overrightarrow{j}_{conv}^{\left( n\right) } &=&{Re}\left\{ \Psi _{n}^{\dag }%
\frac{1}{i\hbar }[\overrightarrow{r},\widehat{H}]\Psi _{n}\right\} \\
\overrightarrow{j}^{\prime \left( n\right) } &=&-\frac{1}{3}{Re}\left\{ \Psi
_{n}^{\dag }(\frac{1}{i\hbar }[\overrightarrow{r},\widehat{H}_{D}]\Psi
_{n})\right\} \\
\overrightarrow{j}_{extra}^{\left( n\right) } &=&\overrightarrow{j}^{\prime
\left( n\right) }+\overrightarrow{K}^{\left( n\right) }.
\end{eqnarray*}%
And%
\begin{eqnarray*}
\overrightarrow{j}_{x}^{\prime \left( n\right) } &=&-\frac{1}{3}{Re}\left\{
\Psi _{n}^{\dag }(\frac{1}{i\hbar }[x,\widehat{H}_{D}]\Psi _{n})\right\} \\
&=&-\frac{1}{6}\{\Psi _{n}^{\dag }(\eta \left( \left(
p_{y}^{2}-p_{z}^{2}\right) \sigma _{x}-2p_{y}p_{x}\sigma
_{y}+2p_{z}p_{x}\sigma _{z}\right) \Psi _{n}) \\
&&+(\eta \left( \left( p_{y}^{2}-p_{z}^{2}\right) \sigma
_{x}-2p_{y}p_{x}\sigma _{y}+2p_{z}p_{x}\sigma _{z}\right) \Psi _{n})^{\dag
}\Psi _{n}\}
\end{eqnarray*}%
So the $x$ component of the extra particle current density is%
\begin{align}
(\overrightarrow{j}_{extra}^{\left( n\right) })_{x}& = & & K_{x}+%
\overrightarrow{j}_{x}^{\prime \left( n\right) }  \notag \\
& = & & \frac{1}{3}\hbar ^{2}\eta \{\left( \nabla _{y}\sigma _{x}\Psi
_{n}\right) ^{\dag }\nabla _{y}\Psi _{n}+\left( \nabla _{x}\sigma _{z}\Psi
_{n}\right) ^{\dag }\nabla _{z}\Psi _{n}+\left( \nabla _{z}\Psi _{n}^{\dag
}\right) \left( \nabla _{x}\sigma _{z}\Psi _{n}\right)  \notag \\
& & & -\left( \nabla _{y}\Psi _{n}^{\dag }\right) \left( \nabla _{x}\sigma
_{y}\Psi _{n}\right) -\left( \nabla _{z}\sigma _{x}\Psi _{n}\right) ^{\dag
}\nabla _{z}\Psi _{n}-\left( \nabla _{x}\sigma _{y}\Psi _{n}\right) ^{\dag
}\nabla _{y}\Psi _{n}\}  \notag \\
& & & +\frac{1}{6}\hbar ^{2}\eta \{\Psi _{n}^{\dag }(\left[ \left( \nabla
_{y}^{2}-\nabla _{z}^{2}\right) \sigma _{x}-2\nabla _{y}\nabla _{x}\sigma
_{y}+2\nabla _{z}\nabla _{x}\sigma _{z}\right] \Psi _{n})  \notag \\
& & & +(\left[ \left( \nabla _{y}^{2}-\nabla _{z}^{2}\right) \sigma
_{x}-2\nabla _{y}\nabla _{x}\sigma _{y}+2\nabla _{z}\nabla _{x}\sigma _{z}%
\right] \Psi _{n})^{\dag }\Psi _{n}\}  \notag \\
& = & & \frac{1}{6}\hbar ^{2}\eta \{\Psi _{n}^{\dag }(\nabla _{y}^{2}\sigma
_{x}\Psi _{n})+(\nabla _{y}^{2}\sigma _{x}\Psi _{n})^{\dag }\Psi
_{n}+2\left( \nabla _{y}\sigma _{x}\Psi _{n}\right) ^{\dag }\nabla _{y}\Psi
_{n}  \notag \\
& & & -\Psi _{n}^{\dag }(\nabla _{z}^{2}\sigma _{x}\Psi _{n})-(\nabla
_{z}^{2}\sigma _{x}\Psi _{n})^{\dag }\Psi _{n}-2\left( \nabla _{z}\sigma
_{x}\Psi _{n}\right) ^{\dag }\nabla _{z}\Psi _{n}  \notag \\
& & & -2\nabla _{x}\nabla _{y}[\Psi _{n}^{\dag }(\sigma _{y}\Psi
_{n})]+2\nabla _{x}\nabla _{z}[\Psi _{n}^{\dag }\left( \sigma _{z}\Psi
_{n}\right) ]\}  \notag \\
& = & & \frac{1}{6}\hbar ^{2}\eta \{(\nabla _{y}^{2}-\nabla _{z}^{2})[\Psi
_{n}^{\dag }(\sigma _{x}\Psi _{n})]-2\nabla _{x}\nabla _{y}[\Psi _{n}^{\dag
}(\sigma _{y}\Psi _{n})]  \notag \\
& & & +2\nabla _{x}\nabla _{z}[\Psi _{n}^{\dag }\left( \sigma _{z}\Psi
_{n}\right) ])\}  \tag{A.6}
\end{align}%
And we can obtain other components of $\overrightarrow{j}_{extra}^{\left(
n\right) }$ by cyclic permutation of indices.
\begin{align}
(\overrightarrow{j}_{extra}^{\left( n\right) })_{y}& =\frac{1}{6}\hbar
^{2}\eta \{(\nabla _{z}^{2}-\nabla _{x}^{2})[\Psi _{n}^{\dag }(\sigma
_{y}\Psi _{n})]~~~~~~~~~~~~~~~  \notag \\
& -2\nabla _{y}\nabla _{z}[\Psi _{n}^{\dag }(\sigma _{z}\Psi _{n})]+2\nabla
_{y}\nabla _{x}[\Psi _{n}^{\dag }\left( \sigma _{x}\Psi _{n}\right) ])\}
\tag{A.7}
\end{align}%
\begin{align}
(\overrightarrow{j}_{extra}^{\left( n\right) })_{z}& =\frac{1}{6}\hbar
^{2}\eta \{(\nabla _{x}^{2}-\nabla _{y}^{2})[\Psi _{n}^{\dag }(\sigma
_{z}\Psi _{n})]~~~~~~~~~~~~~~~  \notag \\
& -2\nabla _{z}\nabla _{x}[\Psi _{n}^{\dag }(\sigma _{x}\Psi _{n})]+2\nabla
_{z}\nabla _{y}[\Psi _{n}^{\dag }\left( \sigma _{y}\Psi _{n}\right) ])\}
\tag{A.8}
\end{align}%
which is the extra term of the current density $\overrightarrow{j}_{extra}$
in Dresselhaus Hamiltonian as stated in the body of the text. It is not zero
in general since $\Psi _{n}^{\dag }(\overrightarrow{\sigma }\Psi _{n})$ is
real and position dependent for general Hamiltonian. Therefore the terms in
above equations are real and not zero after the derivation of position in
general.

\subsection{Extra term of particle current for $\widehat{H}_{I}=\protect%
\beta \widehat{p}^{4}$\ system}

\begin{eqnarray}
\widehat{H} &=&\widehat{H}_{0}+\widehat{H}_{I}+V(r),  \notag \\
\widehat{H}_{I} &=&\beta \widehat{p}^{4}.  \notag
\end{eqnarray}%
\begin{equation}
\frac{1}{i\hbar }[\overrightarrow{r},\widehat{H}_{I}]=4\beta \widehat{p}^{3},
\notag
\end{equation}%
\qquad
\begin{eqnarray*}
\frac{\partial }{\partial t}[\Psi _{n}^{\dag }(\overrightarrow{r})\Psi _{n}(%
\overrightarrow{r})] &=&-{Re}\nabla \cdot \{\Psi _{n}^{\dagger }\frac{1}{%
i\hbar }[\overrightarrow{r},\widehat{H}_{0}]\Psi _{n}\} \\
&&-\frac{1}{4}\Psi _{n}^{\dagger }\left( \nabla \cdot \frac{1}{i\hbar }[%
\overrightarrow{r},\widehat{H}_{I}]\Psi _{n}\right) -\frac{1}{4}(\nabla
\cdot \frac{1}{i\hbar }[\overrightarrow{r},\widehat{H}_{I}]\Psi
_{n})^{\dagger }\Psi _{n} \\
&=&-{Re}\nabla \cdot \{\Psi _{n}^{\dag }\frac{1}{i\hbar }[\overrightarrow{r},%
\widehat{H}_{0}]\Psi _{n}\}-\frac{1}{4}\nabla \cdot \{\Psi _{n}^{\dag }\frac{%
1}{i\hbar }[\overrightarrow{r},\widehat{H}_{I}]\Psi _{n}\} \\
&&-\frac{1}{4}\nabla \cdot \{\left( \frac{1}{i\hbar }[\overrightarrow{r},%
\widehat{H}_{I}]\Psi _{n}\right) ^{\dag }\Psi _{n}\}+\frac{1}{4}\left(
\nabla \Psi _{n}^{\dag }\right) \cdot \{\frac{1}{i\hbar }[\overrightarrow{r},%
\widehat{H}_{I}]\Psi _{n}(\overrightarrow{r})\} \\
&&+\frac{1}{4}\{[\frac{1}{i\hbar }[\overrightarrow{r},\widehat{H}_{I}]\Psi
_{n}(\overrightarrow{r})]^{\dag }\}\cdot \left( \nabla \Psi _{n}\right)
\end{eqnarray*}%
And since%
\begin{eqnarray*}
&&\frac{1}{4}\left( \nabla \Psi _{n}^{\dag }\right) \cdot \frac{1}{i\hbar }[%
\overrightarrow{r},\widehat{H}_{I}]\Psi _{n}+\frac{1}{4}[\frac{1}{i\hbar }[%
\overrightarrow{r},\widehat{H}_{I}]\Psi _{n}]^{\dag }\cdot \left( \nabla
\Psi _{n}\right)  \\
&=&\beta \left\{ \left( \nabla \Psi _{n}^{\dag }\right) \cdot \widehat{p}%
^{3}\Psi _{n}+(\widehat{p}^{3}\Psi _{n}(\overrightarrow{r}))^{\dag }\cdot
\left( \nabla \Psi _{n}\right) \right\}  \\
&=&\beta \left( -i\hbar \right) ^{3}\left\{ \left( \nabla \Psi _{n}^{\dag
}\right) \cdot \left( \nabla ^{3}\Psi _{n}\right) -\left( \nabla ^{3}\Psi
_{n}\right) ^{\dag }\cdot \left( \nabla \Psi _{n}\right) \right\}  \\
&=&\beta \left( -i\hbar \right) ^{3}\nabla \cdot \left\{ \left( \nabla \Psi
_{n}^{\dag }\right) \left( \nabla ^{2}\Psi _{n}\right) -\left( \nabla
^{2}\Psi _{n}\right) ^{\dag }\left( \nabla \Psi _{n}\right) \right\}  \\
&&-\beta \left( -i\hbar \right) ^{3}\left\{ \left( \nabla ^{2}\Psi
_{n}^{\dag }\right) \left( \nabla ^{2}\Psi _{n}\right) -\left( \nabla
^{2}\Psi _{n}\right) ^{\dag }\left( \nabla ^{2}\Psi _{n}\right) \right\}  \\
&=&\beta \left( -i\hbar \right) ^{3}\nabla \cdot \left\{ \left( \nabla \Psi
_{n}^{\dag }\right) \left( \nabla ^{2}\Psi _{n}\right) -\left( \nabla
^{2}\Psi _{n}\right) ^{\dag }\left( \nabla \Psi _{n}\right) \right\} ,
\end{eqnarray*}%
we have%
\begin{eqnarray*}
&&\frac{\partial }{\partial t}[\Psi _{n}^{\dag }(\overrightarrow{r})\Psi
_{n}(\overrightarrow{r})] \\
&=&-\nabla \cdot Re\{\Psi _{n}^{\dag }\frac{1}{i\hbar }[\overrightarrow{r},%
\widehat{H}_{0}]\Psi _{n}\} \\
&&-\nabla \cdot \frac{1}{2}Re\{\Psi _{n}^{\dag }\frac{1}{i\hbar }[%
\overrightarrow{r},\widehat{H}_{I}]\Psi _{n}\} \\
&&+\nabla \cdot \beta \left( -i\hbar \right) ^{3}\left\{ \left( \nabla \Psi
_{n}^{\dag }\right) \left( \nabla ^{2}\Psi _{n}\right) -\left( \nabla
^{2}\Psi _{n}\right) ^{\dag }\left( \nabla \Psi _{n}\right) \right\}
\end{eqnarray*}%
So we get the conserved current%
\begin{eqnarray*}
\overrightarrow{j} &=&Re\{\Psi _{n}^{\dag }\frac{1}{i\hbar }[\overrightarrow{%
r},\widehat{H}_{0}]\Psi _{n}\} \\
&&+\frac{1}{2}Re\{\Psi _{n}^{\dag }\frac{1}{i\hbar }[\overrightarrow{r},%
\widehat{H}_{I}]\Psi _{n}\} \\
&&-\beta \left( -i\hbar \right) ^{3}\left\{ \left( \nabla \Psi _{n}^{\dag
}\right) \left( \nabla ^{2}\Psi _{n}\right) -\left( \nabla ^{2}\Psi
_{n}\right) ^{\dag }\left( \nabla \Psi _{n}\right) \right\}
\end{eqnarray*}%
Compare with the conventional definition,%
\begin{eqnarray*}
\overrightarrow{j}_{conv} &=&Re\{\Psi _{n}^{\dag }\frac{1}{i\hbar }[%
\overrightarrow{r},\widehat{H}_{0}]\Psi _{n}\} \\
&&+Re\{\Psi _{n}^{\dag }\frac{1}{i\hbar }[\overrightarrow{r},\widehat{H}%
_{I}]\Psi _{n}\}
\end{eqnarray*}%
we find the extra term of particle current appears.%
\begin{eqnarray*}
\overrightarrow{j}_{extra} &=&\overrightarrow{j}-\overrightarrow{j}_{conv}=-%
\frac{1}{2}Re\{\Psi _{n}^{\dagger }\frac{1}{i\hbar }[\overrightarrow{r},%
\widehat{H}_{I}]\Psi _{n}\} \\
&&-\beta \left( -i\hbar \right) ^{3}\left\{ \left( \nabla \Psi _{n}^{\dag
}\right) \left( \nabla ^{2}\Psi _{n}\right) -\left( \nabla ^{2}\Psi
_{n}\right) ^{\dag }\left( \nabla \Psi _{n}\right) \right\}
\end{eqnarray*}%
\begin{eqnarray*}
&=&-\beta \left( -i\hbar \right) ^{3}\{\Psi _{n}^{\dag }\nabla \nabla
^{2}\Psi _{n}+\left( \nabla \Psi _{n}^{\dag }\right) \left( \nabla ^{2}\Psi
_{n}\right)  \\
&&-\left( \nabla ^{2}\Psi _{n}\right) ^{\dag }\left( \nabla \Psi _{n}\right)
-\left( \nabla \nabla ^{2}\Psi _{n}\right) ^{\dag }\Psi _{n}\} \\
&=&2\beta \hbar ^{3}{Im}\{\nabla \left( \Psi _{n}^{\dag }\nabla ^{2}\Psi
_{n}\right) \}
\end{eqnarray*}%
And in general case, the state $\Psi _{n}$ can be expressed as%
\begin{equation*}
\Psi _{n}(\overrightarrow{r},t)=\sum_{\overrightarrow{k}_{n}}A(%
\overrightarrow{k}_{n},t)e^{i\overrightarrow{k}_{n}\cdot \overrightarrow{r}}
\end{equation*}%
\begin{eqnarray*}
\overrightarrow{j}_{extra} &=&-\beta i\hbar ^{3}\{\Psi _{n}^{\dag }\nabla
\nabla ^{2}\Psi _{n}+\left( \nabla \Psi _{n}^{\dag }\right) \left( \nabla
^{2}\Psi _{n}\right) -\left( \nabla ^{2}\Psi _{n}\right) ^{\dag }\left(
\nabla \Psi _{n}\right) -\left( \nabla \nabla ^{2}\Psi _{n}\right) ^{\dag
}\Psi _{n}\} \\
&=&-\beta i\hbar ^{3}\sum_{\overrightarrow{k}_{n}}\sum_{\overrightarrow{%
k^{\prime }}_{n}}A^{\ast }(\overrightarrow{k^{\prime }}_{n},t)A(%
\overrightarrow{k}_{n},t)[ik_{n}^{2}\overrightarrow{k}_{n}-ik_{n}^{2}%
\overrightarrow{k_{n}^{\prime }}-ik_{n}^{\prime }{}^{2}\overrightarrow{k}%
_{n}+ik_{n}^{\prime }{}^{2}\overrightarrow{k_{n}^{\prime }}]e^{-i(%
\overrightarrow{k}_{n}-\overrightarrow{k_{n}^{\prime }})\cdot
\overrightarrow{r}} \\
&=&\beta \hbar ^{3}\sum_{\overrightarrow{k}_{n}}\sum_{\overrightarrow{%
k^{\prime }}_{n}}A^{\ast }(\overrightarrow{k^{\prime }}_{n},t)A(%
\overrightarrow{k}_{n},t)[\left( k_{n}^{2}-k_{n}^{\prime }{}^{2}\right) (%
\overrightarrow{k}_{n}-\overrightarrow{k_{n}^{\prime }})]e^{-i(%
\overrightarrow{k}_{n}-\overrightarrow{k_{n}^{\prime }})\cdot
\overrightarrow{r}} \\
&\neq &0
\end{eqnarray*}%
We reached our conclusion that, in general case, $\Psi _{n}(\overrightarrow{r%
},t)$ includes many components $\{A(\overrightarrow{k}_{n},t)e^{i%
\overrightarrow{k}_{n}\cdot \overrightarrow{r}}\},$ therefore the extra
charge current density $\overrightarrow{j}_{extra}$ is position dependent
and not zero.

\subsection{Extra term of total angular momentum current for Rashba system}

First, we have Rashba Hamiltonian: $\widehat{H}=\widehat{H}_{0}+\widehat{H}%
_{R},$ and it is easy to check following relations:%
\begin{eqnarray*}
\frac{1}{i\hbar }[\overrightarrow{r},\widehat{H}_{0}] &=&\frac{1}{m^{\ast }}%
\overrightarrow{p} \\
\frac{1}{i\hbar }[\overrightarrow{r},\widehat{H}_{R}] &=&\alpha \sigma _{y}%
\overrightarrow{e}_{x}-\alpha \sigma _{x}\overrightarrow{e}_{y} \\
\frac{1}{i\hbar }\widehat{H}_{0} &=&-\frac{1}{2}\nabla \cdot \left( \frac{1}{%
i\hbar }[r,\widehat{H}_{0}]\right) \\
\frac{1}{i\hbar }\widehat{H}_{R} &=&-\nabla \cdot \left( \frac{1}{i\hbar }[r,%
\widehat{H}_{R}]\right)
\end{eqnarray*}%
The time evolution of the density of TAM \bigskip $\widehat{J}^{z}$
satisfies following equation%
\begin{eqnarray*}
\frac{\partial }{\partial t}(\Psi _{n}^{\dagger }\widehat{J}^{z}\Psi _{n})
&=&\Psi _{n}^{\dagger }\widehat{J}^{z}\frac{1}{i\hbar }\widehat{H}\Psi _{n}-%
\frac{1}{i\hbar }(\widehat{H}\Psi _{n})^{\dagger }\widehat{J}^{z}\Psi _{n} \\
&=&\frac{1}{i\hbar }\left( \Psi _{n}^{\dagger }(\widehat{H}_{0}+\widehat{H}%
_{R})\widehat{J}^{z}\Psi _{n}\right) -\frac{1}{i\hbar }\left\{ (\widehat{H}%
_{0}+\widehat{H}_{R})\Psi _{n}\right\} ^{\dagger }\widehat{J}^{z}\Psi _{n} \\
&=&-\frac{1}{2}\Psi _{n}^{\dagger }\nabla \cdot \left( \frac{1}{i\hbar }[%
\overrightarrow{r},\widehat{H}_{0}]\widehat{J}^{z}\Psi _{n}\right) -\Psi
_{n}^{\dagger }\nabla \cdot \left( \frac{1}{i\hbar }[\overrightarrow{r},%
\widehat{H}_{R}]\widehat{J}^{z}\Psi _{n}\right) \\
&&-\frac{1}{2}\left( \nabla \cdot \left[ \frac{1}{i\hbar }[r,\widehat{H}%
_{0}]\Psi _{n}\right] ^{\dagger }\right) \widehat{J}^{z}\Psi _{n}-\left(
\nabla \cdot \left[ \frac{1}{i\hbar }[\overrightarrow{r},\widehat{H}%
_{R}]\Psi _{n}\right] ^{\dagger }\right) \widehat{J}^{z}\Psi _{n}
\end{eqnarray*}%
where, in the second step, we used the fact that $\widehat{J}^{z}$ commutes
with the Hamiltonian $\widehat{H}.$
\begin{eqnarray*}
\partial (\Psi _{n}^{\dagger }\widehat{J}^{z}\Psi _{n})/\partial t &=&-\frac{%
1}{2}\nabla \cdot \left( \Psi _{n}^{\dagger }\frac{1}{i\hbar }[%
\overrightarrow{r},\widehat{H}_{0}]\widehat{J}^{z}\Psi _{n}\right) -\frac{1}{%
2}\nabla \cdot \left( \left[ \frac{1}{i\hbar }[\overrightarrow{r},\widehat{H}%
_{0}]\Psi _{n}\right] ^{\dagger }\widehat{J}^{z}\Psi _{n}\right) \\
&&-\nabla \cdot \left[ \Psi _{n}^{\dagger }\left( \frac{1}{i\hbar }[%
\overrightarrow{r},\widehat{H}_{R}]\widehat{J}^{z}\Psi _{n}\right) \right]
-\nabla \cdot \left[ \left( \frac{1}{i\hbar }[\overrightarrow{r},\widehat{H}%
_{R}]\Psi _{n}\right) ^{\dagger }\widehat{J}^{z}\Psi _{n}\right] \\
&&+\frac{1}{2}\left( \nabla \Psi _{n}\right) ^{\dagger }\cdot \left( \frac{1%
}{i\hbar }[\overrightarrow{r},\widehat{H}_{0}]\widehat{J}^{z}\Psi
_{n}\right) +\frac{1}{2}\left( \frac{1}{i\hbar }[\overrightarrow{r},\widehat{%
H}_{0}]\Psi _{n}\right) ^{\dagger }\cdot \nabla \left( \widehat{J}^{z}\Psi
_{n}\right) \\
&&+\left( \nabla \Psi _{n}\right) ^{\dagger }\cdot \left( \frac{1}{i\hbar }[%
\overrightarrow{r},\widehat{H}_{R}]\widehat{J}^{z}\Psi _{n}\right) +\left[
\frac{1}{i\hbar }[\overrightarrow{r},\widehat{H}_{R}]\Psi _{n}\right]
^{\dagger }\cdot \nabla \left( \widehat{J}^{z}\Psi _{n}\right)
\end{eqnarray*}%
Considering%
\begin{eqnarray*}
&&\frac{1}{2}\left( \nabla \Psi _{n}\right) ^{\dagger }\cdot \left( \frac{1}{%
i\hbar }[\overrightarrow{r},\widehat{H}_{0}]\widehat{J}^{z}\Psi _{n}\right) +%
\frac{1}{2}\left( \frac{1}{i\hbar }[\overrightarrow{r},\widehat{H}_{0}]\Psi
_{n}\right) ^{\dagger }\cdot \nabla \left( \widehat{J}^{z}\Psi _{n}\right) \\
&=&\frac{1}{2}\left( \nabla \Psi _{n}\right) ^{\dagger }\cdot \left( \frac{%
\hbar }{im^{\dagger }}\nabla \widehat{J}^{z}\Psi _{n}\right) +\frac{1}{2}%
\left( \frac{\hbar }{im^{\dagger }}\nabla \Psi _{n}\right) ^{\dagger }\cdot
\nabla \left( \widehat{J}^{z}\Psi _{n}\right) =0
\end{eqnarray*}%
\begin{eqnarray*}
&&\left( \nabla \Psi _{n}\right) ^{\dagger }\cdot \left( \frac{1}{i\hbar }[%
\overrightarrow{r},\widehat{H}_{R}]\widehat{J}^{z}\Psi _{n}\right) +\left[
\frac{1}{i\hbar }[\overrightarrow{r},\widehat{H}_{R}]\Psi _{n}\right]
^{\dagger }\cdot \nabla \left( \widehat{J}^{z}\Psi _{n}\right) \\
&=&\left( \nabla \Psi _{n}\right) ^{\dagger }\cdot \left[ \left( \alpha
\sigma _{y}\overrightarrow{e}_{x}-\alpha \sigma _{x}\overrightarrow{e}%
_{y}\right) \widehat{J}^{z}\Psi _{n}\right] +\left[ \left( \alpha \sigma _{y}%
\overrightarrow{e}_{x}-\alpha \sigma _{x}\overrightarrow{e}_{y}\right) \Psi
_{n}\right] ^{\dagger }\cdot \nabla \left( \widehat{J}^{z}\Psi _{n}\right) \\
&=&\nabla \cdot \left\{ \Psi _{n}^{\dagger }\left( \left( \alpha \sigma _{y}%
\overrightarrow{e}_{x}-\alpha \sigma _{x}\overrightarrow{e}_{y}\right)
\widehat{J}^{z}\Psi _{n}\right) \right\} =\nabla \cdot \left\{ \Psi
_{n}^{\dagger }\left( \frac{1}{i\hbar }[\overrightarrow{r},\widehat{H}_{R}]%
\widehat{J}^{z}\Psi _{n}\right) \right\}
\end{eqnarray*}%
Where we have used the following relation%
\begin{equation*}
\left[ \left( \alpha \sigma _{y}\overrightarrow{e}_{x}-\alpha \sigma _{x}%
\overrightarrow{e}_{y}\right) \cdot \nabla \Psi _{n}\right] ^{\dagger
}\left( \widehat{J}^{z}\Psi _{n}\right) =\left( \nabla \Psi _{n}\right)
^{\dagger }\cdot \left( \left( \alpha \sigma _{y}\overrightarrow{e}%
_{x}-\alpha \sigma _{x}\overrightarrow{e}_{y}\right) \widehat{J}^{z}\Psi
_{n}\right)
\end{equation*}%
Therefore, we obtain
\begin{eqnarray*}
\partial (\Psi _{n}^{\dagger }(\overrightarrow{r},t)\widehat{J}^{z}\Psi
_{n})/\partial t &=&-\frac{1}{2}\nabla \cdot \left( \Psi _{n}^{\dagger }%
\frac{1}{i\hbar }[\overrightarrow{r},\widehat{H}_{0}]\widehat{J}^{z}\Psi
_{n}\right) -\frac{1}{2}\nabla \cdot \left( \left[ \frac{1}{i\hbar }[%
\overrightarrow{r},\widehat{H}_{0}]\Psi _{n}\right] ^{\dagger }\widehat{J}%
^{z}\Psi _{n}\right) \\
&&-\nabla \cdot \left[ \Psi _{n}^{\dagger }\left( \frac{1}{i\hbar }[%
\overrightarrow{r},\widehat{H}_{R}]\widehat{J}^{z}\Psi _{n}\right) \right]
-\nabla \cdot \left[ \left( \frac{1}{i\hbar }[\overrightarrow{r},\widehat{H}%
_{R}]\Psi _{n}\right) ^{\dagger }\widehat{J}^{z}\Psi _{n}\right] \\
&&+\nabla \cdot \left\{ \Psi _{n}^{\dagger }\left( \frac{1}{i\hbar }[%
\overrightarrow{r},\widehat{H}_{R}]\widehat{J}^{z}\Psi _{n}\right) \right\}
\end{eqnarray*}%
Since $\frac{1}{i\hbar }[\overrightarrow{r},\widehat{H}_{R}]=\alpha \sigma
_{y}\overrightarrow{e}_{x}-\alpha \sigma _{x}\overrightarrow{e}_{y},$ it
yields
\begin{equation*}
\nabla \cdot \left\{ \Psi _{n}^{\dagger }\left( \frac{1}{i\hbar }[%
\overrightarrow{r},\widehat{H}_{R}]\widehat{J}^{z}\Psi _{n}\right) \right\}
=\nabla \cdot \left\{ \left( \frac{1}{i\hbar }[\overrightarrow{r},\widehat{H}%
_{R}]\Psi _{n}\right) ^{\dagger }\widehat{J}^{z}\Psi _{n}\right\}
\end{equation*}%
Thus, we have%
\begin{eqnarray*}
\partial (\Psi _{n}^{\dagger }\widehat{J}^{z}\Psi _{n})/\partial t &=&-\frac{%
1}{2}\nabla \cdot \left( \Psi _{n}^{\dagger }\frac{1}{i\hbar }[%
\overrightarrow{r},\widehat{H}_{0}]\widehat{J}^{z}\Psi _{n}\right) -\frac{1}{%
2}\nabla \cdot \left( \left[ \frac{1}{i\hbar }[\overrightarrow{r},\widehat{H}%
_{0}]\Psi _{n}\right] ^{\dagger }\widehat{J}^{z}\Psi _{n}\right) \\
&&-\frac{1}{2}\nabla \cdot \left[ \Psi _{n}^{\dagger }\frac{1}{i\hbar }[%
\overrightarrow{r},\widehat{H}_{R}]\widehat{J}^{z}\Psi _{n}\right] -\frac{1}{%
2}\nabla \cdot \left[ \left( \frac{1}{i\hbar }[\overrightarrow{r},\widehat{H}%
_{R}]\Psi _{n}\right) ^{\dagger }\widehat{J}^{z}\Psi _{n}\right] \\
&=&-\frac{1}{2}\nabla \cdot \left( \Psi _{n}^{\dagger }\frac{1}{i\hbar }[%
\overrightarrow{r},\widehat{H}]\widehat{J}^{z}\Psi _{n}\right) -\frac{1}{2}%
\nabla \cdot \left( \left[ \frac{1}{i\hbar }[\overrightarrow{r},\widehat{H}%
]\Psi _{n}\right] ^{\dagger }\widehat{J}^{z}\Psi _{n}\right)
\end{eqnarray*}%
Meanwhile\ \ \ \ \ \ \ \ \ \ \ \ \ \ \ \
\begin{eqnarray*}
\left( \frac{1}{i\hbar }[\overrightarrow{r},\widehat{H}]\Psi _{n}\right)
^{\dagger }\widehat{J}^{z}\Psi _{n} &=&\left[ \left( \frac{1}{m^{\ast }}%
\overrightarrow{p}+\alpha \sigma _{y}\overrightarrow{e}_{x}-\alpha \sigma
_{x}\overrightarrow{e}_{y}\right) \Psi _{n}\right] ^{\dagger }\widehat{J}%
^{z}\Psi _{n} \\
&=&\left[ \left( \frac{\hbar }{im^{\ast }}\nabla +\alpha \sigma _{y}%
\overrightarrow{e}_{x}-\alpha \sigma _{x}\overrightarrow{e}_{y}\right) \Psi
_{n}\right] ^{\dagger }\widehat{J}^{z}\Psi _{n} \\
&=&-\frac{\hbar }{im^{\ast }}\nabla \{\Psi _{n}^{\dagger }\widehat{J}%
^{z}\Psi _{n}\}+\frac{\hbar }{im^{\ast }}\Psi _{n}^{\dagger }(%
\overrightarrow{r},t)\nabla \widehat{J}^{z}\Psi _{n} \\
&&+\Psi _{n}^{\dagger }\left( \alpha \sigma _{y}\overrightarrow{e}%
_{x}-\alpha \sigma _{x}\overrightarrow{e}_{y}\right) \widehat{J}^{z}\Psi _{n}
\\
&=&-\frac{\hbar }{im^{\ast }}\nabla \{\Psi _{n}^{\dagger }\widehat{J}%
^{z}\Psi _{n}\}+\Psi _{n}^{\dagger }(\overrightarrow{r},t)\frac{1}{i\hbar }[%
\overrightarrow{r},\widehat{H}]\widehat{J}^{z}\Psi _{n}
\end{eqnarray*}%
It results%
\begin{align}
\partial (\Psi _{n}^{\dagger }\widehat{J}^{z}\Psi _{n})/\partial t& =-\frac{1%
}{2}\nabla \cdot \left\{ \left( \Psi _{n}^{\dagger }\frac{1}{i\hbar }[%
\overrightarrow{r},\widehat{H}]\widehat{J}^{z}\Psi _{n}\right) +\left( \left[
\frac{1}{i\hbar }[\overrightarrow{r},\widehat{H}]\Psi _{n}\right] ^{\dagger }%
\widehat{J}^{z}\Psi _{n}\right) \right\}  \notag \\
& =-\nabla \cdot \left\{ \Psi _{n}^{\dagger }\frac{1}{i\hbar }[%
\overrightarrow{r},\widehat{H}]\widehat{J}^{z}\Psi _{n}\right\} +\nabla
\cdot \left\{ \frac{\hbar }{2im^{\ast }}\nabla \{\Psi _{n}^{\dagger }%
\widehat{J}^{z}\Psi _{n}\}\right\}  \tag{C.1}
\end{align}%
\begin{equation}
\frac{\partial }{\partial t}\sum_{n}\rho _{n}w_{A}^{(n)}(\overrightarrow{r}%
,t)=\frac{\partial }{\partial t}{Re}\left\{ \sum_{n}\rho _{n}\Psi
_{n}^{\dagger }\widehat{J}^{z}\Psi _{n}\right\}  \notag
\end{equation}%
\begin{equation}
=-\nabla \cdot {Re}\sum_{n}\rho _{n}\left\{ \Psi _{n}^{\dagger }\frac{1}{%
i\hbar }[\overrightarrow{r},\widehat{H}]\widehat{J}^{z}\Psi _{n}-\frac{\hbar
}{2im^{\ast }}\nabla \{\Psi _{n}^{\dagger }\widehat{J}^{z}\Psi
_{n}\}\right\} .  \tag{C.2}
\end{equation}%
And use the fact that $\Psi _{n}^{\dagger }\sigma _{i}\Psi _{n}^{\dagger }$
is real, we have
\begin{eqnarray*}
&&-Re\left\{ \frac{\hbar }{2im^{\ast }}\nabla (\Psi _{n}^{\dagger }\widehat{J%
}^{z}\Psi _{n})\right\} \\
&=&-Re\left\{ \frac{\hbar }{2im^{\ast }}\nabla (\Psi _{n}^{\dagger }\widehat{%
s}^{z}\Psi _{n})\right\} -Re\left\{ \frac{\hbar }{2im^{\ast }}\nabla (\Psi
_{n}^{\dagger }\widehat{l}^{z}\Psi _{n})\right\} \\
&=&-Re\left\{ \frac{\hbar }{2im^{\ast }}\nabla (\Psi _{n}^{\dagger }\widehat{%
l}^{z}\Psi _{n})\right\} \\
&=&-\frac{1}{2}\left\{ \frac{\hbar }{2im^{\ast }}\nabla (\Psi _{n}^{\dagger
}\left( xp_{y}-yp_{x}\right) \Psi _{n})-\frac{\hbar }{2im^{\ast }}\nabla
(\left( xp_{y}-yp_{x}\right) \Psi _{n})^{\dag }\Psi _{n}\right\} \\
&=&\frac{1}{2}\frac{\hbar ^{2}}{2m^{\ast }}\{(\nabla \Psi _{n}^{\dagger
})\left( x\nabla _{y}-y\nabla _{x}\right) \Psi _{n}+(\left( x\nabla
_{y}-y\nabla _{x}\right) \Psi _{n})^{\dag }\nabla \Psi _{n} \\
&&+\Psi _{n}^{\dagger }\left( x\nabla _{y}-y\nabla _{x}\right) \nabla \Psi
_{n}+(\left( x\nabla _{y}-y\nabla _{x}\right) \nabla \Psi _{n})^{\dag }\Psi
_{n}\} \\
&&+\frac{1}{2}\frac{\hbar ^{2}}{2m^{\ast }}\left\{ \Psi _{n}^{\dagger
}\left( e_{x}\nabla _{y}-e_{y}\nabla _{x}\right) \Psi _{n}+(\left(
e_{x}\nabla _{y}-e_{y}\nabla _{x}\right) \Psi _{n})^{\dag }\Psi _{n}\right\}
\end{eqnarray*}%
Therefore,
\begin{align}
\overrightarrow{j}_{L}& ={Re}\sum_{n}\rho _{n}\left\{ \Psi _{n}^{\dagger }(%
\overrightarrow{r},t)\frac{1}{i\hbar }[\overrightarrow{r},\widehat{H}]%
\widehat{J}^{z}\Psi _{n}\right\}  \notag \\
& +\frac{1}{2}\frac{\hbar ^{2}}{2m^{\ast }}\sum_{n}\rho _{n}\left\{ \Psi
_{n}^{\dagger }\left( e_{x}\nabla _{y}-e_{y}\nabla _{x}\right) \Psi
_{n}+(\left( e_{x}\nabla _{y}-e_{y}\nabla _{x}\right) \Psi _{n})^{\dag }\Psi
_{n}\right\}  \notag \\
& +\frac{1}{2}\frac{\hbar ^{2}}{2m^{\ast }}\sum_{n}\rho _{n}\{(\nabla \Psi
_{n}^{\dagger })\left( x\nabla _{y}-y\nabla _{x}\right) \Psi _{n}+(\left(
x\nabla _{y}-y\nabla _{x}\right) \Psi _{n})^{\dag }\nabla \Psi _{n}  \notag
\\
& +\Psi _{n}^{\dagger }\left( x\nabla _{y}-y\nabla _{x}\right) \nabla \Psi
_{n}+(\left( x\nabla _{y}-y\nabla _{x}\right) \nabla \Psi _{n})^{\dag }\Psi
_{n}\}  \tag{C.3}
\end{align}%
The conventional current of TAM $J_{L}^{z}$:%
\begin{equation}
\overrightarrow{j}_{conv}=\frac{1}{2}{Re}\sum_{n}\rho _{n}\Psi _{n}^{\dagger
}\left\{ \frac{1}{i\hbar }[\overrightarrow{r},\widehat{H}]\widehat{J}^{z}+%
\widehat{J}^{z}\frac{1}{i\hbar }[\overrightarrow{r},\widehat{H}]\right\}
\Psi _{n}  \tag{C.4}
\end{equation}%
Due to the following relation for Rashba Hamiltonian:

\begin{equation}
\left[ J_{z},\frac{1}{i\hbar }[\overrightarrow{r},\widehat{H}]\right]
=i\hbar \lbrack (p_{y}/m^{\ast }-\alpha \sigma _{x})\overrightarrow{e}%
_{x}-(p_{x}/m^{\ast }+\alpha \sigma _{y})\overrightarrow{e}_{y}]  \tag{C.5}
\end{equation}%
we have%
\begin{align}
\overrightarrow{j}_{conv}& =\frac{1}{2}{Re}\sum_{n}\rho _{n}\Psi
_{n}^{\dagger }\left\{ \frac{1}{i\hbar }[\overrightarrow{r},\widehat{H}]%
\widehat{J}^{z}+\widehat{J}^{z}\frac{1}{i\hbar }[\overrightarrow{r},\widehat{%
H}]\right\} \Psi _{n}  \notag \\
& ={Re}\left\{ \sum_{n}\rho _{n}\Psi _{n}^{\dagger }\frac{1}{i\hbar }[%
\overrightarrow{r},\widehat{H}]\widehat{J}^{z}\Psi _{n}\right\}  \notag \\
& +\frac{1}{2}{Re}\sum_{n}\rho _{n}\left\{ i\hbar \Psi _{n}^{\dagger
}[(p_{y}/m^{\ast })\overrightarrow{e}_{x}-(p_{x}/m^{\ast })\overrightarrow{e}%
_{y}]\Psi _{n}\right\}  \notag \\
& ={Re}\left\{ \sum_{n}\rho _{n}\Psi _{n}^{\dagger }\frac{1}{i\hbar }[%
\overrightarrow{r},\widehat{H}]\widehat{J}^{z}\Psi _{n}\right\}  \notag \\
& +\frac{1}{4}\frac{\hbar ^{2}}{m^{\ast }}\left\{ \Psi _{n}^{\dagger
}(\nabla _{y}\overrightarrow{e}_{x}-\nabla _{x}\overrightarrow{e}_{y})\Psi
_{n}+\left( (\nabla _{y}\overrightarrow{e}_{x}-\nabla _{x}\overrightarrow{e}%
_{y})\Psi _{n}\right) ^{\dag }\Psi _{n}\right\}  \tag{C.6}
\end{align}%
where, in the second step, the property that $\Psi _{n}^{\dagger }\sigma
_{i}\Psi _{n}^{\dagger }$ is real is again used. Compare (C.6) with (C.3),
we have%
\begin{eqnarray*}
\overrightarrow{j}_{extra}^{\left( n\right) } &=&\overrightarrow{j}%
_{L}^{\left( n\right) }-\overrightarrow{j}_{conv}^{\left( n\right) } \\
&=&\frac{1}{2}\frac{\hbar ^{2}}{2m^{\ast }}\{(\nabla \Psi _{n}^{\dagger
})\left( x\nabla _{y}-y\nabla _{x}\right) \Psi _{n}+(\left( x\nabla
_{y}-y\nabla _{x}\right) \Psi _{n})^{\dag }\nabla \Psi _{n} \\
&&+\Psi _{n}^{\dagger }\left( x\nabla _{y}-y\nabla _{x}\right) \nabla \Psi
_{n}+(\left( x\nabla _{y}-y\nabla _{x}\right) \nabla \Psi _{n})^{\dag }\Psi
_{n}\}
\end{eqnarray*}%
So the components of $\overrightarrow{j}_{extra}^{\left( n\right) }$ are
\begin{eqnarray*}
j_{extra}^{\left( n\right) x} &=&\frac{1}{2}\frac{\hbar ^{2}}{2m^{\ast }}%
\{(\nabla _{x}\Psi _{n}^{\dagger })\left( x\nabla _{y}-y\nabla _{x}\right)
\Psi _{n}+(\left( x\nabla _{y}-y\nabla _{x}\right) \Psi _{n})^{\dag }\nabla
_{x}\Psi _{n} \\
&&+\Psi _{n}^{\dagger }\left( x\nabla _{y}-y\nabla _{x}\right) \nabla
_{x}\Psi _{n}+(\left( x\nabla _{y}-y\nabla _{x}\right) \nabla _{x}\Psi
_{n})^{\dag }\Psi _{n}\} \\
&=&\frac{\hbar ^{2}}{4m^{\ast }}[x(\nabla _{x}\Psi _{n}^{\dagger })\nabla
_{y}\Psi _{n}+x(\nabla _{y}\Psi _{n}^{\dagger })\nabla _{x}\Psi _{n}+x\Psi
_{n}^{\dagger }\nabla _{y}\nabla _{x}\Psi _{n} \\
&&+x(\nabla _{x}\nabla _{y}\Psi _{n}^{\dagger })\Psi _{n}-y\nabla
_{x}^{2}(\Psi _{n}^{\dagger }\Psi _{n})] \\
&=&\frac{\hbar ^{2}}{4m^{\ast }}[x\nabla _{x}\nabla _{y}(\Psi _{n}^{\dagger
}\Psi _{n})-y\nabla _{x}^{2}(\Psi _{n}^{\dagger }\Psi _{n})] \\
&=&\frac{\hbar ^{2}}{4m^{\ast }}[x\nabla _{y}-y\nabla _{x}]\nabla _{x}(\Psi
_{n}^{\dagger }\Psi _{n}) \\
&=&-\frac{1}{4m^{\ast }}l^{z}p_{x}(\Psi _{n}^{\dagger }\Psi _{n})
\end{eqnarray*}%
Similarly, we have%
\begin{equation}
j_{extra}^{\left( n\right) y}=\frac{\hbar ^{2}}{4m^{\ast }}[x\nabla
_{y}-y\nabla _{x}]\nabla _{y}(\Psi _{n}^{\dagger }\Psi _{n})  \tag{C.7} \\
=-\frac{1}{4m^{\ast }}l^{z}p_{y}(\Psi _{n}^{\dagger }\Psi _{n})  \notag
\end{equation}%
So finally we arrive at%
\begin{equation}
\overrightarrow{j}_{extra}=-\frac{1}{4m^{\ast }}\sum_{n}\rho _{n}l^{z}%
\overrightarrow{p}(\Psi _{n}^{\dagger }\Psi _{n})  \tag{C.8}
\end{equation}%
Where $\Psi _{n}^{\dagger }\Psi _{n}\ $is the density of particles and real.
It must be position dependent in general, thus $\overrightarrow{j}_{extra}=%
\frac{\hbar ^{2}}{4m^{\ast }}[x\nabla _{y}-y\nabla _{x}]\nabla (\Psi
_{n}^{\dagger }\Psi _{n})\neq 0$.

\end{document}